\documentclass[10pt,journal,compsoc]{IEEEtran}
\pdfoutput=1

\usepackage[T1]{fontenc}
\usepackage{amsmath,amssymb,amsthm}
\usepackage{array}
\usepackage{booktabs}
\usepackage{cite}
\usepackage{graphicx}
\usepackage{pgfplots}
\usepackage{tikz}
\usetikzlibrary{arrows.meta,fit,positioning,shapes.geometric,shapes.symbols}
\pgfplotsset{compat=1.18}
\usepackage[expansion=false]{microtype}
\usepackage{xcolor}
\usepackage{xspace}
\usepackage{url}
\usepackage[hidelinks]{hyperref}
\usepackage{pdfpages}

\newcolumntype{L}[1]{>{\raggedright\arraybackslash}p{#1}}
\providecommand{\IEEEPARstart}[2]{#1#2}
\makeatletter
\@ifundefined{IEEEkeywords}{%
  \newenvironment{IEEEkeywords}{\par\small\noindent\textbf{Index Terms---}}{\par}
}{}
\makeatother

\newcommand{\systemname}{FactGate}
\newcommand{\system}{FactGate\xspace}
\newcommand{\code}[1]{\texttt{\detokenize{#1}}}

\newcommand{\facts}{\mathcal{F}}
\newcommand{\accounted}{\mathcal{K}}
\newcommand{\effect}{\mathcal{E}}
\newcommand{\rieffect}{\mathcal{E}^{ri}}

\newtheorem{definition}{Definition}
\newtheorem{property}{Property}
\newtheorem{theorem}{Theorem}
\newtheorem{proposition}{Proposition}

\newcommand{\ExposureCorpusHash}{\texttt{5704d22606b0}}
\newcommand{\RQOneCases}{24}
\newcommand{\RQOnePassed}{24}
\newcommand{\RQOneRows}{16}

\newcommand{\RQThreeCases}{5}

\newcommand{\RQFourVFourGatesPassed}{30}
\newcommand{\RQFourVFourGatesTotal}{30}

\newcommand{\RQFourVFourOracleFacts}{1,035,013}

\newcommand{\RQTwoExposureCases}{19}

\newcommand{\RQTwoExposureMismatches}{0}

\newcommand{\ExposureFormalStates}{410,766}

\newcommand{\BitmapFormalStates}{122,976}

\newcommand{\OutcomeFormalStates}{20}

\newcommand{\ArtifactFormalStates}{1,324}

\newcommand{\VFiveFamilyPropertySeeds}{24}

\newcommand{\VFiveFamilyPropertyMaxTasks}{6}
\newcommand{\VFiveFamilyPropertyQueries}{718}

\newcommand{\VFiveFamilyPropertyRefused}{465}

\newcommand{\VFiveFamilyPropertyNarrowedTasks}{67}

\newcommand{\RQFiveTransitions}{3}
\newcommand{\RQFiveChecksPassed}{5}
\newcommand{\RQFiveChecksTotal}{5}

\newcommand{\RQFiveBytesPerFact}{620}
\newcommand{\RQFiveBuildMicrosPerFact}{22.8}

\newcommand{\FinalVFiveProvSQLLeafSamples}{3}

\newcommand{\FinalVFiveProvSQLLeafTotalRowFacts}{336,003}

\newcommand{\FinalVFiveCounterExactMaxDep}{12}
\newcommand{\FinalVFiveCounterNaiveMinDep}{18}

\newcommand{\FinalVFiveScaleSevenLargestDependency}{10,500,000}

\newcommand{\FinalVFiveAdversaryOwnerBits}{6}
\newcommand{\FinalVFiveAdversaryTightenedBits}{4}

\newcommand{\FinalVFiveSubphaseGoSidePctSOneSFOne}{57}

\newcommand{\FinalVFiveSubphaseSamplesSTwoSFTen}{1}
\newcommand{\FinalVFiveSubphaseMicrosPerFactSTwoSFTen}{2.0}

\newcommand{\FinalVFiveSubphaseGoSidePctSSixHundredKByteSixteen}{86}

\newcommand{\FinalVFiveSubphaseMicrosPerFactSSixHundredKByteSixteen}{0.7}

\newcommand{\FinalVFivePublicationCampaign}{\texttt{formal-v113-publication-05}}
\newcommand{\FinalVFivePublicationCommit}{\texttt{faf4d095701b}}
\newcommand{\FinalVFivePublicationDeployments}{33}
\newcommand{\FinalVFivePublicationProfiles}{11}
\newcommand{\FinalVFivePublicationFreshExecutions}{3}
\newcommand{\FinalVFivePublicationProfileCells}{125}
\newcommand{\FinalVFivePublicationScaleNonProfileCells}{36}
\newcommand{\FinalVFivePublicationCompilerNonProfileCells}{11}
\newcommand{\FinalVFivePublicationTotalCells}{172}
\newcommand{\FinalVFivePublicationMeasuredSamples}{9,393}
\newcommand{\FinalVFivePublicationBaselineCells}{58}

\newcommand{\FinalVFivePublicationBaselinePerCell}{90}
\newcommand{\FinalVFivePublicationReplayZeroSQLSamples}{2,340}
\newcommand{\FinalVFivePublicationOverheadMin}{21.8}
\newcommand{\FinalVFivePublicationOverheadMax}{144.0}
\newcommand{\FinalVFivePublicationSOneNovelS}{0.33}
\newcommand{\FinalVFivePublicationSOneReleaseFacts}{100,000}
\newcommand{\FinalVFivePublicationSTwoNovelS}{2.46}
\newcommand{\FinalVFivePublicationSTwoRows}{3}
\newcommand{\FinalVFivePublicationSTwoDependencyFacts}{1,200,000}
\newcommand{\FinalVFivePublicationBaselineTableBody}{%
S1/SF1 & 5,000 & 10,000 & 20,000 & 2.91 & 100.51 & 61.80 & 27.44 & 34.5$\times$ \\
S1/SF10 & 50,000 & 100,000 & 200,000 & 12.62 & 326.93 & 108.95 & 49.76 & 25.9$\times$ \\
S2/SF1 & 3 & 9 & 120,000 & 14.14 & 384.32 & 62.28 & 24.37 & 27.2$\times$ \\
S2/SF10 & 3 & 9 & 1,200,000 & 30.42 & 2459.32 & 62.65 & 24.12 & 80.8$\times$ \\
S3/10k-50k & 10,000 & 10,000 & 50,000 & 2.43 & 151.06 & 62.34 & 28.65 & 62.1$\times$ \\
S3/1k-5k & 1,000 & 1,000 & 5,000 & 0.83 & 97.15 & 56.61 & 25.01 & 116.9$\times$ \\
S3/45k-225k & 45,000 & 45,000 & 225,000 & 8.67 & 319.42 & 88.91 & 43.67 & 36.9$\times$ \\
S4/depth-4 & 3 & 9 & 120,000 & 13.89 & 429.24 & 72.95 & -- & 30.9$\times$ \\
S5/SF1 & 5,000 & 10,000 & 20,000 & 4.61 & 224.39 & 65.60 & 26.20 & 48.7$\times$ \\
S5/SF10 & 50,000 & 100,000 & 200,000 & 19.52 & 1431.39 & 112.37 & 50.19 & 73.3$\times$ \\
S6/100k-x16 & 100,000 & 1,600,000 & 1,700,000 & 173.89 & 3797.28 & -- & -- & 21.8$\times$ \\
S6/100k-x4 & 100,000 & 400,000 & 500,000 & 51.14 & 1133.42 & -- & -- & 22.2$\times$ \\
S6/100x16 & 100 & 1,600 & 1,700 & 0.90 & 129.82 & -- & -- & 144.0$\times$ \\
S6/100x4 & 100 & 400 & 500 & 0.76 & 85.20 & -- & -- & 112.0$\times$ \\
S6/10k-x16 & 10,000 & 160,000 & 170,000 & 16.28 & 586.97 & -- & -- & 36.0$\times$ \\
S6/10k-x4 & 10,000 & 40,000 & 50,000 & 5.33 & 201.72 & -- & -- & 37.8$\times$ \\%
}
\newcommand{\FinalVFivePublicationPhaseDominantSOneSFOne}{execute\_and\_derive}
\newcommand{\FinalVFivePublicationPhaseDominantSharePctSOneSFOne}{32.3}
\newcommand{\FinalVFivePublicationPhaseSettlementMSSOneSFOne}{20.59}

\newcommand{\FinalVFivePublicationPhaseDominantSSixHundredKByteSixteen}{execute\_and\_derive}
\newcommand{\FinalVFivePublicationPhaseDominantSharePctSSixHundredKByteSixteen}{77.2}
\newcommand{\FinalVFivePublicationPhaseSettlementMSSSixHundredKByteSixteen}{296.95}

\newcommand{\FinalVFivePublicationNovelSpreadMaxPct}{5.1}
\newcommand{\FinalVFivePublicationDirectSpreadMaxPct}{10.5}
\newcommand{\FinalVFivePublicationConcurrencySpreadMaxPct}{7.0}
\newcommand{\FinalVFivePublicationConcurrencySerialDrainPFiftyMS}{133.4}
\newcommand{\FinalVFivePublicationConcurrencyFiftyNaturalDrainPFiftyMS}{4642.0}
\newcommand{\FinalVFivePublicationConcurrencyFiftyNaturalDrainPNinetyFiveMS}{5353.3}
\newcommand{\FinalVFivePublicationConcurrencyFiftyNaturalRequestsPerSecond}{10.8}
\newcommand{\FinalVFivePublicationConcurrencyTenNaturalRequestsPerSecond}{15.8}

\newcommand{\FinalVFivePublicationProvSQLFortyFiveKPostgreSQLMS}{47}
\newcommand{\FinalVFivePublicationProvSQLFortyFiveKProvSQLS}{90.9}
\newcommand{\FinalVFivePublicationProvSQLFortyFiveKBDGS}{2.62}
\newcommand{\FinalVFivePublicationProvSQLFortyFiveKDependencyFacts}{1,305,003}

\newcommand{\FinalVFivePublicationProvSQLDependencyLinksVerified}{270}

\newcommand{\FinalVFivePublicationAttackSamples}{135}

\newcommand{\FinalVFivePublicationAttackSequences}{7}
\newcommand{\FinalVFivePublicationAttackCorpusDigest}{\texttt{483b667369a0}}

\newcommand{\FinalVFivePublicationAttackPaginationReleaseFacts}{12}
\newcommand{\FinalVFivePublicationAttackPaginationDependencyFacts}{18}

\newcommand{\FinalVFivePublicationAttackThresholdCeiling}{5}

\newcommand{\FinalVFivePublicationAttackThresholdRejectionCode}{\texttt{EXPOSURE\_BUDGET\_EXHAUSTED}}

\newcommand{\FinalVFivePublicationAttackThresholdProbes}{3}
\newcommand{\FinalVFivePublicationAttackThresholdAnswered}{2}

\newcommand{\FinalVFivePublicationRLSPrefixRelease}{(1,1) (2,2) (3,3) (4,4) (5,5) (6,6) (7,6) (8,6) (9,6) (10,6) (11,6) (12,6) (13,6) (14,6) (15,6) (16,6) (17,6) (18,6) (19,6) (20,6) (21,6) (22,6) (23,6) (24,6) (25,6) (26,6) (27,6) (28,6) (29,6) (30,6) (31,6) (32,6) (33,6) (34,6) (35,6) (36,6) (37,6) (38,6) (39,6) (40,6) (41,6) (42,6) (43,6) (44,6) (45,6) (46,6) (47,6) (48,6) (49,6) (50,6) (51,6) (52,6) (53,6) (54,6) (55,6) (56,6) (57,6) (58,6) (59,6) (60,6) (61,6) (62,6) (63,6) (64,6) (65,6) (66,6) (67,7) (68,8) (69,9) (70,9) (71,9) (72,9) (73,9) (74,9) (75,9) (76,9) (77,9) (78,9) (79,9) (80,9) (81,9) (82,9) (83,9) (84,9) (85,9) (86,9) (87,9) (88,9) (89,9) (90,9) (91,9) (92,9) (93,9) (94,9) (95,9) (96,9) (97,10) (98,10) (99,10) (100,10)}
\newcommand{\FinalVFivePublicationRLSBudgetRelease}{7}
\newcommand{\FinalVFivePublicationRLSUnionRelease}{10}
\newcommand{\FinalVFivePublicationRLSPrefixDependency}{(1,2) (2,4) (3,6) (4,8) (5,10) (6,12) (7,12) (8,12) (9,12) (10,12) (11,12) (12,12) (13,12) (14,12) (15,12) (16,12) (17,12) (18,12) (19,12) (20,12) (21,12) (22,12) (23,12) (24,12) (25,12) (26,12) (27,12) (28,12) (29,12) (30,12) (31,12) (32,12) (33,12) (34,12) (35,12) (36,12) (37,18) (38,18) (39,18) (40,18) (41,18) (42,18) (43,18) (44,18) (45,18) (46,18) (47,18) (48,18) (49,18) (50,18) (51,18) (52,18) (53,18) (54,18) (55,18) (56,18) (57,18) (58,18) (59,18) (60,18) (61,18) (62,18) (63,18) (64,18) (65,18) (66,18) (67,18) (68,18) (69,18) (70,18) (71,18) (72,18) (73,18) (74,18) (75,18) (76,18) (77,18) (78,18) (79,18) (80,18) (81,18) (82,18) (83,18) (84,18) (85,18) (86,18) (87,18) (88,18) (89,18) (90,18) (91,18) (92,18) (93,18) (94,18) (95,18) (96,18) (97,18) (98,18) (99,18) (100,18)}
\newcommand{\FinalVFivePublicationRLSBudgetDependency}{12}
\newcommand{\FinalVFivePublicationRLSUnionDependency}{18}
\newcommand{\FinalVFivePublicationRLSPrefixOutcome}{(1,2) (2,4) (3,6) (4,8) (5,10) (6,12) (7,13) (8,14) (9,15) (10,15) (11,15) (12,15) (13,15) (14,15) (15,15) (16,15) (17,15) (18,15) (19,15) (20,15) (21,15) (22,15) (23,15) (24,15) (25,15) (26,15) (27,15) (28,15) (29,15) (30,15) (31,15) (32,15) (33,15) (34,15) (35,15) (36,15) (37,17) (38,17) (39,19) (40,19) (41,21) (42,21) (43,21) (44,21) (45,21) (46,21) (47,21) (48,21) (49,21) (50,21) (51,21) (52,21) (53,21) (54,21) (55,21) (56,21) (57,21) (58,21) (59,21) (60,21) (61,21) (62,21) (63,21) (64,21) (65,21) (66,21) (67,23) (68,24) (69,26) (70,26) (71,26) (72,26) (73,26) (74,26) (75,26) (76,26) (77,26) (78,26) (79,26) (80,26) (81,26) (82,26) (83,26) (84,26) (85,26) (86,26) (87,26) (88,26) (89,26) (90,26) (91,26) (92,26) (93,26) (94,26) (95,26) (96,26) (97,27) (98,27) (99,27) (100,27)}
\newcommand{\FinalVFivePublicationRLSBudgetOutcome}{18}
\newcommand{\FinalVFivePublicationRLSUnionOutcome}{27}

\newcommand{\FinalVFivePublicationRLSPerCell}{9}
\newcommand{\FinalVFivePublicationRLSQueries}{100}
\newcommand{\FinalVFivePublicationRLSRowsRLS}{250}

\newcommand{\FinalVFivePublicationRLSBoundedSuccessful}{36}
\newcommand{\FinalVFivePublicationRLSBoundedRefusalIndex}{37}
\newcommand{\FinalVFivePublicationRLSBoundedLedger}{6/12/15}

\newcommand{\FinalVFivePublicationRLSDrainRLSMS}{16}
\newcommand{\FinalVFivePublicationRLSDrainUnlimitedS}{7.30}

\newcommand{\FinalVFivePublicationScaleSamples}{2,160}
\newcommand{\FinalVFivePublicationScaleDependencyLinksVerified}{7,560}
\newcommand{\FinalVFivePublicationScaleDependencyLinkMaxFacts}{2,070,000}
\newcommand{\FinalVFivePublicationScaleCells}{24}
\newcommand{\FinalVFivePublicationScaleSettleTenKMS}{40}
\newcommand{\FinalVFivePublicationScaleSettleMillionMS}{163}
\newcommand{\FinalVFivePublicationScaleSettleMillionFullOverlapMS}{124}
\newcommand{\FinalVFivePublicationScaleMillionHistory}{1,035,000}
\newcommand{\FinalVFivePublicationScaleMillionDrainMS}{1,306}

\newcommand{\FinalVFivePublicationIndependentOracleSamples}{2,214}
\newcommand{\TPCHQueries}{22}

\newcommand{\TPCHLowerableExplicitJoin}{0}

\newcommand{\TPCHSubqueryQueries}{7}
\newcommand{\TPCHArithmeticQueries}{5}

\newcommand{\SSBQueries}{13}

\newcommand{\SSBLowerableExplicitJoin}{2}

\newcommand{\AgentWorkloadQueries}{40}
\newcommand{\AgentWorkloadLowerable}{28}
\newcommand{\AgentWorkloadLowerablePct}{70}

\newcommand{\GeneratedPlans}{5,000}
\newcommand{\GeneratedFixtures}{50}

\newcommand{\GeneratedMismatches}{0}
\newcommand{\GeneratedHashMismatches}{0}
\newcommand{\GeneratedReleaseFacts}{55,801}
\newcommand{\GeneratedDependencyFacts}{202,306}

\newcommand{\GeneratedSplitChecks}{380}

\newcommand{\HeldoutQueries}{40}
\newcommand{\HeldoutLowerable}{22}

\newcommand{\HeldoutLowerablePct}{55}

\newcommand{\HeldoutEngineeringGaps}{11}
\newcommand{\HeldoutNewRule}{6}
\newcommand{\HeldoutAgentDefects}{1}

\newcommand{\SweepBenignAuthorized}{23}

\newcommand{\SweepRecoverAt}{2}

\newcommand{\TimingPooledSDMS}{2.49}

\newcommand{\TimingFullMin}{2}
\newcommand{\TimingFullMax}{18}

\newcommand{\AlignedRowsNaturalAdmitted}{59}
\newcommand{\AlignedRowsNaturalExecuted}{58}

\newcommand{\AgentRuns}{36}

\newcommand{\ThroughputOneRootFiftyDisjointNPS}{29}
\newcommand{\ThroughputFourRootFiftyDisjointNPS}{66}

\newcommand{\ThroughputOneRootFiftyRefused}{76}
\newcommand{\ThroughputOneRootFiftyRequests}{600}

\newcommand{\RetryK}{3}

\newcommand{\RetryOneRootFiftyCompletionPct}{100.0}

\newcommand{\RetryOneRootFiftyDisjointEndToEndPNinetyFive}{1656}

\newcommand{\BFiveSOneCommonPct}{53}
\newcommand{\BFiveSSixCommonPct}{22}

\newcommand{\BFiveSSixArmIV}{3217}
\newcommand{\BFiveSSixDerivation}{2313}
\newcommand{\BFiveSSixLedger}{72}

\newcommand{\NaiveScaleMillionZeroRatio}{75}

\newcommand{\NaiveScaleMillionFullRatio}{38}

\newcommand{\NaiveScaleTenKZeroRatio}{2.0}
\newcommand{\NaiveScaleTenKFullRatio}{0.7}

\newcommand{\SheetCorrectedRange}{21--23}

\newcommand{\RawEvidenceFiles}{216}
\newcommand{\RawEvidenceMB}{614}
\newcommand{\RawEvidenceSealedSamples}{48}

\newcommand{\RawEvidenceDOI}{10.5281/zenodo.23149437}

\title{\systemname: A Task-Scoped Data Gateway for AI Agents with
Cumulative Exposure Accounting}

\author{Minmin~Wu%
\IEEEcompsocitemizethanks{\IEEEcompsocthanksitem M.~Wu is with China Telecom Global Limited.\protect\\
E-mail: \mbox{wuminmin@chinatelecomglobal.com}.}}
\date{}

\begin{document}
\bstctlcite{IEEEexample:BSTcontrol}
\maketitle

\begin{abstract}
Per-query authorization does not compose across an adaptive agent's pages,
retries, rewrites, and delegated subtasks, and row or query counters are
order-dependent. \system is a release layer for immutable reporting
publications that turns one approved task into a cumulative contract: three
set-valued budgets over declared identities (result, positive-output
dependency, and outcome facts) shared by the root task and descendants.
Exact set novelty against the root-family ledger is settled atomically on one
PostgreSQL snapshot before any result becomes retrievable. The guarantee is bounded
\emph{release}: within one root family and publication epoch, at most
the approved number of distinct declared facts of each kind becomes
available, whatever the admitted sequence; inference from answered or refused
predicates is not bounded. Accounting is exact under declared rules over
a closed fragment (connected
equijoins, grouping, bounded unions, nested Views, derived-cell arithmetic)
that admits \HeldoutLowerablePct\% of \HeldoutQueries{} held-out LLM-written
statements (\AgentWorkloadLowerablePct\% of
a development set adapted during construction). Ordinals, bitmaps, and a
Merkle-radix set keep novelty exact at $10^7$ facts. A
single-host campaign (\FinalVFivePublicationMeasuredSamples{} samples)
charges every scripted decomposition once, keeps an adaptive trace within
its budgets, and costs
\FinalVFivePublicationOverheadMin{}--\FinalVFivePublicationOverheadMax{}$\times$
a direct query; settlement takes a median
\FinalVFivePublicationScaleSettleMillionMS{}~ms against a million-fact
history at zero overlap.
\end{abstract}


\section{Introduction}
\label{sec:introduction}

\IEEEPARstart{D}{atabase} authorization asks, ``Can this query execute?''
An adaptive agent adds, ``How much policy-defined exposure has this task
acquired?'' Within one approved task, an agent may split projections,
overlap pages, retry, rewrite, and delegate (each child grant
human-approved) while individual queries remain legal. Per-query
authorization and row limits do not state the root family's distinct
cumulative exposure. \system accounts that contract-defined quantity, not the
agent's knowledge. Data-flow control (DFC) asks whether a SQL flow is
allowed \cite{summers2026dataflow,summers2026passant}; given
authorization and any DFC check, \system asks whether the next allowed execution
fits the root's cumulative contract.

Query count ignores answer size, row count makes a one-row aggregate over
millions of source rows look cheap, and bytes depend on encoding. Query
pricing develops anti-arbitrage objectives
\cite{koutris2012query,deep2017qirana,peng2025aria} and provenance explains
where results came from \cite{green2007provenance,amsterdamer2011aggregate};
\system asks how human authorization can create a stateful release
budget conserved across adaptive querying and delegation. Enterprise
operational analytics commonly serves scheduled ETL outputs rather than
OLTP primaries. \system compiles each completed cut into an immutable
publication; new roots may bind the next cut, and unfinished roots stay on
the prior one through version routing or fail closed. It therefore bounds the
governed release of declared facts within one reporting epoch, not
freshness over a changing database.

Large enterprises operate lakes or warehouses, reporting marts, Data
Catalogs, identity directories, and approval workflows, but none turns one
approved purpose into an enforceable contract across an agent's queries,
retries, pages, and delegated subtasks. Marketplace-integrated agents close the discovery-and-request half of this
gap (Data Product MCP
\cite{tonnarelli2026dataproductmcp,dietz2025automating,bitol2025odcs}), but
after approval, per-query checks still do not state how much
governed data the task has released. \system starts at that boundary
as a brownfield-compatible egress layer at the governed serving tier.
The reporting pipeline supplies immutable publications, the Catalog
supplies Products and attested keys, and identity/workflow
systems supply subject, purpose, and approval. \system compiles these into a
signed task contract at the database boundary and never exposes the raw lake
to the agent.

\system budgets the root task in three set-valued dimensions. \emph{Accounted
result exposure} identifies selected result values (wire label \emph{release});
accounting can precede availability. The \emph{positive-output dependency
footprint} identifies the base facts the declared rules read---neither minimal
causes nor all physical reads---so a one-cell
aggregate can expose large input. \emph{Query outcome} combines canonical
predicate atoms with one normalized query/result composite. Existing members
add no novelty: related thresholds reuse shared atoms but charge different
literals/composites; exact replay is free.

The system follows a withhold-before-settlement discipline
(\code{reserve -> replay lookup / execute+stream -> derive -> settle -> publish};
Fig.~\ref{fig:architecture-settlement}). The visible query and ordinal
companion share one frozen PostgreSQL snapshot and remain withheld while one
Control transaction computes exact set novelty, checks all limits, advances
the root head, and commits audit, receipt, and a PENDING intent. A recoverable
object-store step then promotes the staged object, and a second transaction
records AVAILABLE/CONSUMED.

\begin{figure*}[t]
\centering
\resizebox{0.78\textwidth}{!}{%
\begin{tikzpicture}[
  font=\scriptsize,
  box/.style={draw, rounded corners=1.5pt, align=center, minimum height=6mm,
    inner xsep=3pt, fill=blue!5},
  store/.style={draw, cylinder, shape border rotate=90, aspect=.25,
    align=center, minimum height=9mm, minimum width=16mm, fill=green!7},
  txnbox/.style={draw, rounded corners=1.5pt, align=center, minimum width=18mm,
    minimum height=6mm, inner xsep=2pt, fill=blue!5},
  decision/.style={draw, diamond, aspect=2.0, align=center, inner sep=1.5pt,
    fill=yellow!12},
  arrow/.style={-{Latex[length=2mm]}, semithick},
  exchange/.style={{Latex[length=2mm]}-{Latex[length=2mm]}, semithick},
  return/.style={-{Latex[length=2mm]}, semithick, dashed}
]
% (a) Offline publication
\node[font=\footnotesize\bfseries] at (2.6,4.3) {(a) Offline immutable publication};
\node[box] (snapshot) at (1.2,2.8) {frozen business\\snapshot + Catalog};
\node[box, minimum width=20mm] (compiler) at (1.2,1.6)
  {Snapshot Index\\Compiler};
\node[box, fill=green!7, minimum width=22mm] (sidecar) at (4.0,2.8)
  {PG sidecar\\entity $\to$ handle};
\node[box, fill=green!7, minimum width=22mm] (dictionary) at (1.2,.45)
  {FactID $\leftrightarrow$ ordinal\\hot hash / cold payload};
\node[box, fill=green!10, minimum width=22mm] (manifest) at (4.0,1.6)
  {Catalog publication\\binds all digests};
\draw[arrow] (snapshot) -- (compiler);
\draw[arrow] (compiler) -- (sidecar);
\draw[arrow] (compiler) -- (dictionary);
\draw[arrow] (dictionary) -- (manifest);
\draw[arrow] (sidecar) -- (manifest);

% (b) Online path
\node[font=\footnotesize\bfseries] at (12.5,4.3) {(b) Online exact set settlement};
\node[box, fill=red!5, minimum width=14mm] (client) at (7.35,2.65) {Agent\\plan / result};
\node[decision] (cache) at (10.4,2.65) {authorize +\\semantic cache};
\node[txnbox, minimum width=26mm] (execute) at (13.65,2.65)
  {visible SQL + streamed\\ordinal companion};
\node[txnbox, minimum width=23mm] (derive) at (17.05,2.65)
  {bitmap + atom/\\composite effect};
\node[txnbox, minimum width=25mm] (cached) at (13.65,1.55)
  {committed observation\\reference};
\node[txnbox, minimum width=21mm] (novelty) at (17.05,1.45)
  {exact novelty\\+ union};
\node[txnbox, minimum width=21mm] (cas) at (15.6,.45)
  {three-ledger\\root-head CAS};
\node[txnbox, fill=green!10, minimum width=26mm] (commit) at (12.45,.45)
  {commit ledgers/audit/receipt\\+ PENDING intent};
\node[box, fill=blue!7, minimum width=25mm] (promote) at (7.35,.45)
  {promote + verify\\AVAILABLE/CONSUMED};
% one snapshot: the publication feeds the online visible query
\draw[arrow] (manifest.east) -- (5.4,1.6) -- (5.4,3.5) -- (13.65,3.5) -- (execute.north);
\node[font=\scriptsize, above] at (8.15,3.5) {one snapshot};
\coordinate (snapcorner) at (7.0,3.9);
\draw[arrow] (client) -- node[above, font=\scriptsize]{submit} (cache);
\draw[return] (cache.south west) |- node[pos=.7, below, font=\scriptsize]{refuse} ([xshift=4mm,yshift=-4.5mm]client.south) -- ([xshift=4mm]client.south);
\draw[arrow] (cache) -- node[above, pos=.25]{miss} (execute);
\draw[arrow] (cache) |- node[near start,left]{hit} (cached);
\draw[arrow] (execute) -- (derive);
\draw[arrow] (derive) -- (novelty);
\draw[arrow] (cached) -- (novelty);
\draw[arrow] (novelty.south) |- (cas.east);
\draw[arrow] (cas) -- (commit);
\draw[arrow] (commit) -- (promote);
\draw[return] (promote.north) -- node[pos=.3, left, font=\scriptsize]{result\_id} (promote.north |- client.south);
\node[draw, dashed, rounded corners, inner sep=2mm,
  fit=(cache)(execute)(cached)(derive)(novelty)(cas)(commit)(promote)(snapcorner),
  label={[font=\footnotesize]below:trusted enforcement boundary}] {};
\end{tikzpicture}
}
\caption{(a) Offline publication fixes ordinals and the Merkle-radix Outcome
set; (b) online exact settlement, promotion, and replay.}
\label{fig:architecture-settlement}
\end{figure*}

Beneath the accounting sits a task-scoped authorization substrate: a control
plane that turns an approved task into a signed, budgeted, auditable database
session with safe views, row scope, denied fields, operation limits,
per-session counters, and receipts. Everything that makes the budget a set
rather than a counter is this article's contribution. It is organized
around three difficulties: deriving an admitted query's fact sets exactly (1--2);
deciding which rewrites preserve identity and which create a new Outcome
(3); and keeping retrievable results tied to settled facts under
concurrency, failure, replay, and delegation (3--5). The five
implementation-backed contributions are:

(1) A typed FactID semantics for accounted Result, positive-output
Dependency, and canonical-atom/composite Outcome sets in a restricted
algebra. (2) A same-snapshot ordinal companion for a closed reporting
fragment. The fragment covers one-Product and connected equijoin queries
(two- and three-source joins evaluated end to end; up to 16 roles admitted
by the compiler), two-branch
union, grouping with HAVING and the admitted aggregates (COUNT, COUNT
DISTINCT, SUM, MIN, MAX, exact AVG), and single-Product exact
$\{+,-,\times\}$ projections settled as derived cells; bounded
nested Views flatten into it. It has no scalar functions, division, OR,
subqueries, or outer joins (Section~\ref{sec:limits} measures the
consequence). (3) Profile-relative novelty and atomic three-budget
settlement per root family and publication epoch. Novelty is exact by the
declared rules and verified as consistent with our reference implementation;
it is refined by immutable ordinal dictionaries, compressed bitmaps, and an
exact Merkle-radix Outcome set. Settlement precedes \code{AVAILABLE},
semantic replay has zero novelty, and a rejection leaks a threshold bit
bounded by the request budget. (4) Signed, monotonic delegation whose
descendants share one root-family ledger, so fan-out cannot multiply
release capacity, while per-task request budgets multiply refusal capacity
by the approved tree size. (5) Algebraic
propositions, a normal-form soundness theorem, an abstract settlement-safety
property sanity-checked by two-fact TLA+ models, and executable
Go/PostgreSQL differential, generated-plan, and fault-injection evidence.

\section{Context and Related Work}
\label{sec:related}

\subsection{Closest System Families}

The supplement tabulates each family's object, horizon, subject, and
enforcement/release coupling.

\textbf{History, pricing, and provenance.} Auditing reconstructs access or
determines whether a history fixes forbidden data; simulatable auditing
also protects denial
\cite{agrawal2004hippocratic,motwani2008auditing,kenthapadi2005simulatable}.
Query pricing studies determinacy, history, and information gain
\cite{koutris2012query,deep2017qirana,peng2025aria}, and DP accounting and answer
reuse bound $\epsilon$ \cite{dwork2014foundations,mazmudar2022cachedp}.
DProvDB \cite{zhang2023dprovdb} is the closest shape---a persistent
per-analyst cumulative budget---but meters DP loss over noised views, not
an exact declared-identity set (reproduced side-by-side: supplement).
Statistical-database inference control studied trackers, query-set overlap,
and cumulative auditing
\cite{dobkin1979secure,denning1980tracker,nabar2006robustness};
\system's Dependency ledger is a provenance-derived engineering of that
lineage: it bounds settled predicates, not inference, and its budget is a
set of declared facts where PINQ's is $\epsilon$ \cite{mcsherry2009pinq}. Zero-novelty replay is the
syntactic analogue of history-aware pricing, and
Theorem~\ref{thm:nf-soundness} covers rules R1--R6 only, weaker than
arbitrage-freedom. Why/where/how provenance
\cite{buneman2001why,green2007provenance,cheney2009provenance,amsterdamer2011aggregate}
and the GProM/ProvSQL capture mechanisms
\cite{arab2018gprom,senellart2018provsql,sen2026provsql} explain derivations;
\system adds no provenance algebra but settles cumulative novelty of typed
effects before release.

\textbf{Data-flow policy enforcement.} FlowGuard and DFC/Passant are the
closest comparators \cite{summers2026dataflow,summers2026passant}.
Passant enforces optimizer-invariant aggregate predicates over each output
tuple's provenance monomials (flow shapes \system does not subsume);
trajectory policy is its future work. \system differs in temporal unit: DFC asks whether a flow is
allowed, \system whether another allowed flow fits the cumulative contract
across queries, retries, pages, rewrites, and descendants (composition not
evaluated; Passant artifact replay: supplement).

\textbf{Authorization, delegation, and agent controls.} Task-based and
purpose-based authorization scope rights to an activity or a declared purpose
\cite{thomas1997tbac,byun2008purpose}, and attenuation-only delegation is the
macaroon model \cite{birgisson2014macaroons}. Query-rewriting fine-grained
control \cite{rizvi2004extending} is the per-query comparator for the closed
fragment; usage/history control, RLS/VPD, ABAC, and query-policy systems
establish ongoing authority, quotas, and rows
\cite{park2004ucon,abadi2003history,li2003delegation,postgres16rls,mehta2017qapla,zhang2022blockaid}. MCP authorization
and delegation scope capability invocation
\cite{mcp2025spec,south2025delegation}. AgentDojo evaluates
model behavior under prompt injection \cite{debenedetti2024agentdojo} and
SafeNLIDB inference risk on the model side \cite{liu2026safenlidb}.
Provenance-based access control \cite{park2012pbac} and CaMeL's tool-layer
capabilities \cite{debenedetti2025camel} decide admission, not cumulative
release, and GAAP's persistent cross-tool flow state \cite{stanley2026gaap}
spans steps and services outside \system's scope. \system assumes the authorization boundary, settles every descendant
against one root ledger, treats the agent and its SQL as untrusted, and
claims neither result sanitization, flow tracking, nor inference safety.

\section{Exposure Model}
\label{sec:model}

\textbf{Adversary and goal.} The agent, its plans, retries, pages, rewrites,
and delegated calls are untrusted and may be adaptive. The security goal is
bounded governed release: within one approved root family on one publication
epoch, no more than $B_X$ distinct FactIDs of each kind become
\code{AVAILABLE}, whatever sequence of admitted queries the agent issues.
Four extraction patterns motivate the three budgets, three of them
exercised by the evaluation. Page-by-page sweeps are bounded by $B_R$
(Section~\ref{sec:calibration}); aggregates that read far more base facts
than a purpose warrants by $B_D$ (the adaptive trace); delegation fan-out
by the shared root ledger (the delegated child's refusal). Probe ladders
are bounded by $B_O$ in count but not in recovery: under the default
$B_O=128$ a binary search recovers a 64-bit cell in 64 settled probes, and
$B_Q$ adds 128 refusal bits per task. An executed adversary study realizes
the bound live (retained pilot, supplement): recovery is capped at
$\lfloor B_O/2\rfloor$ bits under owner-derived tiers
(\FinalVFiveAdversaryOwnerBits{}/\FinalVFiveAdversaryTightenedBits{}
bits), exact recovery occurs only under a loosened grant, and greedy
extraction stops exactly at $B_D$. Four
boundaries hold at this goal: it counts declared identities, not inference;
page order is order-only in the Catalog tie-break keys alone (the
\code{ORDER BY} rule below); fresh per-root, per-epoch capacity bounds
cross-day release by the approval workflow, not by \system; and the settlement
decision is two data-dependent counter channels. The first is rejection: a
rejected attempt is a chosen membership question over unaccounted facts
answered by one bit. It is charged against $B_Q$
only and never $B_O$, so rejection capacity is $\sum B_Q$ over the approved
task tree, not $B_X$ (Section~\ref{sec:online}): under the default route
($B_O{=}B_Q{=}128$) a $k$-task tree can settle 128 distinct predicates in
total but take $128k$ refusals. The second is the settled response, which
returns the per-dimension charge and the ledger's used and limit counters
(\code{exposure}, \code{exposure_budget}). Each of at most $B_O$
successful novel queries thus reveals $|\Delta_X|$, up to $3\lceil\log_2
B_X\rceil$ bits of query-set-overlap information across the three
dimensions. A third, timing channel exists because execution precedes
settlement, so a refused query's latency can depend on the footprint it did
not release. Execution costs \FinalVFiveSubphaseMicrosPerFactSTwoSFTen{}--\FinalVFiveSubphaseMicrosPerFactSSixHundredKByteSixteen{}~$\mu$s
per Dependency Fact in the sub-phase pilot; retained refusals with full
footprints of \TimingFullMin{}--\TimingFullMax{} Facts neither establish nor
exclude the dependence against \TimingPooledSDMS{}~ms noise, so the channel
is unquantified and outside the guarantee (supplement).
Proposition~\ref{prop:probing-bound} bounds the response payload only.
Returning only the
remaining budget would be a planning convenience, not a smaller channel
(remaining equals prior remaining minus charge). The model
trusts identity, attribute, and approval assertions; Catalog and
Exposure-Profile authors and the immutable publication pipeline; PostgreSQL
execution and snapshot semantics; enforcement and Control services; key
management; and object-store access control.

\subsection{Publication Epochs and Task Binding}

\system's upstream boundary is a completed reporting synchronization, not a
mutable transactional database.

\begin{definition}[Publication epoch]
A publication epoch $P_e$ binds a datasource, its schema, the snapshot
bundle mapping every stable role to its immutable reporting publication,
the Catalog, and the ordinal dictionaries (descriptor and digest $h_e$ in
the supplement). The signed Catalog SHA transitively commits them. The
upstream synchronization cutoff is a trusted ETL premise, and a new
synchronization creates $P_{e+1}$ rather than mutating $P_e$.
\end{definition}

A deployment alias such as \code{latest publication} resolves to a concrete
Catalog digest at approval time, never at query time (a draft that must
follow a rollover is reissued). Let $\mathsf Q(T)$ contain every query issued by an approved root
$T$ or any delegated descendant. Then $\operatorname{publication}(T)=P_e$
and $\operatorname{publication}(q)=P_e$ for every $q\in\mathsf Q(T)$.
The Catalog transitively binds the grant and FactID environment to that
epoch. Ledger state, replay metadata, evidence, audits, and receipts carry
the same bindings; execution, settlement, and replay fail closed if
they differ, so an active root cannot switch publications.

\subsection{Facts and Task State}

\begin{definition}[Approved task]
\label{def:task}
An approved root task is the immutable authorization boundary
$T=(P,S,B,C)$ with ceilings $B=(B_R,B_D,B_O)$.
Here $P$ is the signed policy grant over the principal, data products,
fields, and scopes, and $S$ is the immutable publication/snapshot binding of
the epoch above. $B$ gives the release, positive-output
dependency-footprint, and query-outcome ceilings together with the request
budget $B_Q$, the maximum number of settled or rejected attempts; a
rejection consumes one unit plus its admission-reserved rows and database
time and charges no exposure set. Finally, $C$ is the signed set of
semantic and execution constraints and row/time guards.  In particular, $C$ binds the effect
profile and closed QueryPlan language, type/collation versions, semantic-View
contracts, delegation rules, replay identity, and result handling.  The
monotone accounted ledger $K(T)$ is mutable protocol state and is deliberately
\emph{not} a component of the approved tuple.
\end{definition}

The release/dependency algebra treats a FactID as a semantic object (its
SHA-256 is only a database index); its semantic environment is
$\Pi=\langle pg,\mathcal T,UTC,\mathcal C,D,h_C\rangle$: a PostgreSQL major
version, admitted type domain, timezone, exact deterministic collation
name/version map, immutable snapshot map, and catalog/lineage digest.  A
judgment is defined only for an \emph{admissible} $\Pi$: source schemas and
entity keys are attested, every collatable field has a deterministic collation,
and volatile expressions are absent.  Let $\varpi$ be the versioned
exposure-profile identifier (Section~\ref{sec:clearance-profile}), $\nu$ a
catalog-owned source namespace (not a view alias), $s$ an immutable snapshot,
$e$ a canonical stable entity key, $a$ a role-qualified field ID, $\tau\in\mathcal T$, and
$C_\tau(v)$ its typed canonical value.  Base identities are
$f_{row}=\langle\varpi,\nu,s,e\rangle$ and
$f_{cell}=\langle\varpi,\nu,s,e,a,\tau,C_\tau(v)\rangle$. A derived identity is
$f_d=\langle\varpi,\mathcal B,k_o,N_\varpi(x),\tau,C_\tau(v),CW(W)\rangle$.
Here $\mathcal B=\operatorname{sort}\{(\nu,s)\}$ is the relation's snapshot
bundle, $k_o$ is the canonical output-row entity key, $N_\varpi(x)$ erases aliases
and normalizes the supported expression, and
$W:\facts_{base}\rightharpoonup\mathbb N^+$
is a finite-support witness multiset (empty support allowed). The
product identity is the namespace projection of $\mathcal B$, not a
caller-supplied name. $CW(W)$ length-prefixes the sorted
$(\operatorname{FactHash}(f),W(f))$ pairs. The durable key is
$\mathrm{SHA256}(d_R\|\operatorname{Canon}(f))$ under the release/dependency
separator $d_R$; Control PostgreSQL stores $\operatorname{Canon}(f)$
and aborts on a same-hash, different-payload row. Outcome facts use
disjoint tagged kinds under separator $d_O$.

Let $N(q)$ be the accepted query normal form, $h_N$ its SHA-256, $m$ the
visible row count, $h_y$ the canonical result-observation commitment, and
$h_c$ the predicate-context commitment to the approved Catalog, immutable
publication/lineage, stable roles, and resolved expressions. Atomization
normalizes each accepted literal occurrence by PostgreSQL type, collation
name/version, operator, and canonical literal value. After exact
deduplication, each semantic atom is
$a_i=\langle k_a,\varpi,h_c,p,r,f,e,\tau,\kappa,op,v\rangle$ with
$k_a=\code{PREDICATE_ATOM}$, $p,r,f,e$ the semantic Product, stable role,
public FieldID, and resolved-expression digest, and $\tau,\kappa,op,v$ the
type, collation, operator, and canonical literal; $A(q)$ is the atom set
$\operatorname{Atomize}(q)$ yields. For any outcome Fact $f$, its physical key is
$\operatorname{FactHash}_O(f)=H(d_O\|\operatorname{Canon}(f))$.
Let $h_A$ commit the sorted $\operatorname{FactHash}_O$ values of $A(q)$ and
$k_c=\code{COMPOSITE_OUTCOME}$. \system creates one semantic composite object
$c_q=\langle k_c,\varpi,\operatorname{profile}(N),h_N,h_y,m,h_c,h_A,|A(q)|\rangle$;
the query's Outcome Facts are $\facts_O(q)=A(q)\cup\{c_q\}$ and $M_O(q)$
is the set of their $\operatorname{FactHash}_O$ values.
Repeated literals and supported rewrites thus reuse atom identities. The
composite commits to the normalized query, exact result observation,
and complete atom set, so every distinct successful query (every page
included) adds one composite, and $B_O$ also caps distinct successful
queries. The Merkle-radix set stores $M_O(q)$; empty results and scalar
zero stay explicit, and the signed grant bounds literal occurrences, unique
atoms, and payloads before settlement.

\begin{proposition}[Fact identity well-definedness]
\label{thm:fact-identity}
For every value in the supported PostgreSQL type domain, canonical encoding
is deterministic and injective over the typed semantic payload (that is, over
PostgreSQL semantic-value equivalence, not raw float NaN bit patterns; proof
in the supplement).
\end{proposition}

\begin{definition}[Exposure effect]
For query $q$ over snapshot $D_s$, the exposure effect
$\effect(q,D_s)=\langle \facts_R,\facts_D,\facts_O\rangle$ contains accounted
result facts (retaining the wire label \code{release})
$\facts_R$ and positive-output dependency-footprint facts $\facts_D$ under a
versioned profile $\varpi$.  The latter are the base facts the inductive rules
below select for successfully derived output rows and cells; the admitted language
requires every \code{ORDER BY} column to be selected, so only the Catalog
tie-break suffix of Page is order-only and outside this profile.
$\facts_O=A(q)\cup\{c_q\}$ contains the predicate footprint and composite
query/result identity defined above.
\end{definition}

\emph{Example (charged by the operator rules of Section~\ref{sec:effect-extraction}).} Over a five-row Product with fields $(id,dept,amt)$, the
query $\pi_{id,amt}\sigma_{dept='S'}$ selects rows $r_1$--$r_3$. It charges
$6$ Result cells, $12$ Dependency facts (each selected row's row fact, its
$dept$ cell read by the predicate, and its two visible cells), and $2$
Outcome facts (the atom $dept{=}\text{'S'}$ and the composite). Splitting it
into $\pi_{id}$ and $\pi_{amt}$ charges $0/0/2$ (two new composites, the same
atom); a page returning $r_2,r_3$ charges $0/0/1$; adding $amt>5$ charges
one new atom and one composite and no new base facts; an exact replay by the
root or by a delegated child charges $0/0/0$.

For a successful $q$, settlement commits the complete candidate Result set
before the later, recoverable publication, so
$F_R^{\mathrm{available}}(q)\subseteq F_R^{\mathrm{accounted}}(q)
=F_R^{\mathrm{candidate}}(q)$; download is untracked, so accounting only
upper-bounds observation.

\begin{definition}[Task-scoped cumulative exposure]
\label{def:task-exposure}
Let $Q=\langle q_1,\ldots,q_n\rangle$ be a finite query-attempt history for a
root and its descendants, and let $Q^+\subseteq Q$ contain exactly the queries
that successfully settle under $T$. For $X\in\{R,D,O\}$,
$\facts_X(q,S)$ abbreviates $\facts_X(q,D_s)$ for the snapshot $D_s$ and
profile that $S$ and $C$ transitively bind; the two spellings denote one set.
The task exposure is
\begin{equation}
 \begin{aligned}
 E(T,Q)&=\langle E_R(T,Q),E_D(T,Q),E_O(T,Q)\rangle,\\
 E_R(T,Q)&=\bigcup_{q\in Q^+}\facts_R(q,S),\\
 E_D(T,Q)&=\bigcup_{q\in Q^+}\facts_D(q,S),\\
 E_O(T,Q)&=\bigcup_{q\in Q^+}\facts_O(q,S).
 \end{aligned}
 \label{eq:task-exposure}
\end{equation}
The components are accounted result exposure (wire label
\code{release}), positive-output dependency
footprint, and query-outcome exposure. Failed or rejected attempts contribute
no set because \system commits none of their effects; the
rejection bit is not accounted here. We write the singleton history as
$E(T,q)$.
\end{definition}

For every root task, the ledger service separately stores previously accounted
sets $K(T)=\accounted_T=\langle\accounted_T^R,\accounted_T^D,
\accounted_T^O\rangle$ and the approved limits
$B=\langle B_R,B_D,B_O\rangle$. Settlement computes

\begin{align}
 \Delta_T^R(q) &= \facts_R(q,D_s)\setminus\accounted_T^R,\\
 \Delta_T^D(q) &= \facts_D(q,D_s)\setminus\accounted_T^D,\\
 \Delta_T^O(q) &= \facts_O(q,D_s)\setminus\accounted_T^O.
 \label{eq:delta}
\end{align}

\system measures cardinality independently in each dimension. A query may settle
only if
$|\accounted_T^R|+|\Delta_T^R|\leq B_R$ and
$|\accounted_T^D|+|\Delta_T^D|\leq B_D$ and
$|\accounted_T^O|+|\Delta_T^O|\leq B_O$.

\subsection{Exposure Profiles as Clearance Manifests}
\label{sec:clearance-profile}

\begin{definition}[Profile-relative clearance exactness]
\label{def:profile-relative-exactness}
For an approved $T=(P,S,B,C)$, $C$ binds one versioned Exposure Profile $\varpi$.
For any admissible $\Pi$ and successful query in the closed language, write
$\facts_X^\varpi(q)$ for the existing component $\facts_X(q,S)$ with its profile
binding made explicit. \system exactly computes the canonical Result, Dependency,
and Outcome sets $\facts_R^\varpi(q),\facts_D^\varpi(q),\facts_O^\varpi(q)$ and their
componentwise novelty
\[
 \Delta_T^X(q)=\facts_X^\varpi(q)\setminus\accounted_T^X,
 \qquad X\in\{R,D,O\}.
\]
This definition fixes the scope of the word \emph{exact}: governance
accounting exactness under the signed contract and closed semantics,
asserting nothing beyond the results that follow, and established
empirically only as consistency with our reference implementations
(Section~\ref{sec:evaluation}). FactIDs are accounting identities, not
estimates of information content, knowledge, or privacy loss; \system prevents
unaccounted governed release under a task-bound, versioned profile.
\end{definition}
\emph{Example (cont.).} Under budgets $B_R/B_D/B_O=6/12/7$ the five-row
episode of Section~\ref{sec:model} ends with ledger totals $(6,12,7)$; a
next query $\sigma_{dept='T'}$ needs new base and Outcome facts, so its
$\Delta_T$ no longer fits, and it is refused uncharged---the ledger still
reads $(6,12,7)$.
Exactness rests on four accounting obligations, proved for the closed
language in the supplement (O1--O4). (O1) Every selected cell of a successful result
has its materialized identity in $\facts_R$. (O2) $\facts_D^\varpi$ is exactly
the declared inductive footprint (retained predicate fields, join keys, group
keys, complete aggregate arguments), excluding rejected rows and non-output
pages. (O3) The listed rewrites, splits, page partitions, overlaps,
retries, join support, and delegation cannot reduce cumulative charge.
(O4) Each success creates one composite and its atoms, exact replay reuses
them, and a rejected attempt creates none, charged only to $B_Q$.

Six obligations separate the assurance layers (O1--O4 semantic claims, O5
models plus transaction tests, O6 binding); none alone proves the
implementation.

\begin{property}[O5: No unaccounted governed release]
\label{prop:no-unaccounted-release}
Let $\mathsf{AVAILABLE}(q)$ mean that $q$'s canonical artifact has reached the
\code{AVAILABLE} state. For every modeled query and $X\in\{R,D,O\}$,
\[
 \mathsf{AVAILABLE}(q)\Longrightarrow
 \facts_X^\varpi(q)\subseteq\accounted_T^X
 \ \wedge\ |\accounted_T^X|\leq B_X.
\]
\end{property}
It is the release-side counterpart of consumable credentials and quotas
\cite{bowers2007consumable,pollack2007quota}.
Execution and private staging are not release; settlement may leave an
artifact \code{PENDING}, so the converse fails and rejection cannot
create \code{AVAILABLE}. Evidence: bounded artifact/ledger models plus
Control-PG fault-injection, receipt, and availability tests, not a
mechanized proof.

\textbf{O6: Profile binding.} $C$ binds effect/normal-form profiles,
type/collation, View contract, replay identity, and result handling;
derivation, settlement, replay, audit, and receipt bind one
task/Catalog/publication/dictionary context, so a changed context cannot
reuse an old replay.

\textbf{Three-way reconciliation.} The effect manifest, root-ledger heads,
and artifact/receipt/availability evidence must identify the same settled
observation and intent; any mismatch fails closed, and neither a receipt
alone nor an availability event without settlement suffices.

\begin{proposition}[Outcome separation and replay]
\label{thm:outcome-separation}
For successful queries in one predicate context, exact replay and supported
equivalent rewrites produce the same atom set and composite identity and hence
zero incremental Outcome charge. Changing one canonical predicate component
changes that atom except with a SHA-256 collision; changing the complete normal
form or result observation changes the composite except with a collision.
Consequently related probes reuse unchanged atoms but charge each novel atom
and novel composite exactly once, including for zero/empty answers (proof in
the supplement).
\end{proposition}

\subsection{Closed Typed Query Language}

Operational multi-source SQL first lowers to
$G_J=\langle V_J,E_J,\Theta_J\rangle$: $2$--$16$ distinct Catalog stable-role
nodes and connected equality edges. Stable-role ordering and a deterministic
fold erase aliases and accepted parenthesization, so the same connected graph
has one core term. Repeated roles, self edges, duplicate equalities, and type or
collation disagreement reject; the operational bounds and construction are
supplemental.

The complete core syntax is
\begin{align*}
e ::= {}& \operatorname{Scan}[\nu,s,r,\Gamma]
 \mid \sigma_{p;F}(e) \mid \pi_A(e) \mid \pi_{A;x}(e) \mid e\bowtie_{\Theta}e \\
 &\mid e\cup_\delta e \mid \gamma_{G;\mathcal A}(e)
 \mid \operatorname{Page}_{O,K,o,l}(e),\\
p ::= {}& b \mid p\wedge b,\qquad b ::= a\;op\;c,\\
op\in {}&\{=,\ne,<,\le,>,\ge\}
 \cup\{\code{LIKE},\code{IN},\code{NOT IN}\},
\end{align*}
where $F$ is exactly the fields read by $p$, $\Theta$ is a nonempty
conjunction of equalities, $\cup_\delta$ is union-distinct,
$\mathcal A\subseteq\{\code{COUNT(*)},\code{COUNT},\code{COUNT DISTINCT},\code{SUM},\code{MIN},
\code{MAX},\code{AVG}\}$ (\code{AVG} exact-numeric only), $\pi_{A;x}$ adds one
single-Product exact $\{+,-,\times\}$ derived cell $x$, and \code{HAVING} is
$\sigma$ over $\gamma$'s output (rules in the supplement).  No implicit SQL construct is hidden in $e$: outer joins,
difference, negation beyond the listed scalar predicates, windows, recursion,
subqueries, volatile functions, and \code{UNION ALL} are not terms.

The partial judgment $\Pi\vdash e:\Gamma$ enforces child-local fields,
role-disjoint typed joins, identical distinctable union schemas, typed
grouping, and a proven total page order; undefined premises reject.

The View front end lowers a bounded acyclic DAG of projections,
inner equijoins, literal conjunctions, and at most one grouping barrier into
this core, rejecting richer SQL or unsafe dependencies; no general View
equivalence is claimed.

Predicates use PostgreSQL three-valued logic and retain only true rows, whereas
grouping/distinct use typed \code{IS NOT DISTINCT FROM}. The closed canonical
domain fixes integer/float widths, exact numeric and temporal encodings,
deterministic collation name/version, \code{jsonb} structural equality, and
PostgreSQL aggregate result types. Non-associative
floating SUM, unsupported casts, and overflow are undefined and settle
nothing; the complete type domain and encoders are supplemental.

\subsection{Annotated-Bag Evaluation and Effect Extraction}
\label{sec:effect-extraction}

A relation is a typed PostgreSQL bag with unique occurrence keys and optional
proven total order; rows and cells carry support $S$ (lineage
\cite{cui2001lineage,buneman2001why}), a positive witness multiset $W$
(how-provenance \cite{green2007provenance}), canonical lineage, and optional
materialized release identity, and every operator preserves
\begin{equation}
 \operatorname{support}(W_r)=S_r,\qquad
 \operatorname{support}(W_a)=S_a.                 \label{eq:anninv}
\end{equation}
Same-proof composition unions support and adds multiplicities; alternative
proofs union support and take pointwise maximum, so alternatives are
idempotent while join fanout remains visible. Row identity
$\operatorname{RID}(R,t)$ hashes the canonical schema, snapshot bundle, and
complete operator-specific key material under a typed, domain-separated
$H$ (construction and collision checks are supplemental).

Evaluation is the partial judgment $\Pi;D\vdash e\Downarrow R$: a failed
premise yields no relation, effect, settlement, or release; every rule
constructs each output cell explicitly (complete rules, NULL/empty-input
cases, and page bounds are in the supplement). Union is bag-aware and idempotent only when its typed tuples are proven
duplicate-free, grouping/distinct equivalence is \code{IS NOT DISTINCT FROM},
floating-point \code{SUM} is rejected because the bag fixes no fold order,
and pagination appends Catalog keys until the order is total; the relation is
a partial function, not a semantics for arbitrary SQL.
Materialization $\operatorname{mat}(R,t,a)$ reuses a cell's pre-existing
release fact when value, type, bundle, and support or expression/witness
still validate, and otherwise mints the derived identity; the effect is
\begin{align}
\facts_R(R,V)&=\{\operatorname{mat}(R,t,a)\mid t\in R,\ a\in V\},\\
\facts_D(R,V)&=\bigcup_{t\in R}S_r(t)\ \cup
                \bigcup_{t\in R,a\in V}S_a(t),\\
\rieffect_\Pi(e,D,V)&=\operatorname{normalize}
  \langle\facts_R(R,V),\facts_D(R,V)\rangle,       \label{eq:observe}
\end{align}
where $S_r(t)$ is row $t$'s support (the base row facts and the predicate,
join-key, group-key, and aggregate-argument cells its derivation read) and
$S_a(t)$ the support of its cell $a$. Query bags and witness multiplicity
are retained during composition, while the two base ledgers are sets.
Equation~\eqref{eq:observe} is the declared
positive-output dependency-footprint profile: it is exact with respect to these
inductive rules, not with respect to arbitrary causal provenance or physical
source access.  Empty answers have an empty base effect except the single
visible row of a global aggregate; \system still adds the tested predicate
atoms and one composite outcome.

\begin{proposition}[Representation closure and dependency determinism]
\label{thm:representation-closure}
For every admissible $\Pi$, typed finite $e$, and immutable $D$, evaluation
has at most one result, satisfying Eq.~\eqref{eq:anninv} with canonical
schema/bundle and unique row keys, valid as input to every well-typed
parent rule; fixed $\Pi$, $e$, $D$, and visible set $V$ uniquely determine
$\facts_D$ and $\rieffect_\Pi(e,D,V)$ whenever evaluation succeeds.
\end{proposition}
\emph{Proofs in the supplement.}

\subsection{Conservation Properties}

The properties below cover only the stated operators and rewrites, not
arbitrary equivalent SQL; the supplement proves Fact-set normalization and the
restricted selection--projection invariance the rewrite tests use.

\begin{definition}[Profile-indexed typed normal form $NF_{\Pi,\varpi}$]
\label{def:nf}
$NF_{\Pi,\varpi}$ maps accepted, well-typed inputs to canonical terms under
the normal-form profile the contract binds (rules R1--R6 below);
equality is syntactic. Its typed digest is the query-semantic component of
replay/outcome identity, distinct from final-ledger Fact-set
$\operatorname{normalize}$.
\end{definition}

\begin{theorem}[Typed normal-form soundness]
\label{thm:nf-soundness}
For admissible $\Pi$, fixed profile $\varpi$, immutable snapshot map $D$, $q_1,q_2$ in the closed
algebra or accepted graph-lowerable SQL fragment, and the same visible field
set $V$,
\[
 NF_{\Pi,\varpi}(q_1)=NF_{\Pi,\varpi}(q_2)\Longrightarrow
 \effect_\Pi(q_1,D,V)=\effect_\Pi(q_2,D,V).
\]
This is not completeness for arbitrary equivalent SQL: normalization neither
invents nor removes transitive/redundant equalities.
\end{theorem}
The theorem is the analogue, for the release/dependency effect and rules
R1--R6, of the invariance of semiring provenance under equivalent
positive-algebra rewrites \cite{green2007provenance,amsterdamer2011aggregate};
what is new is the effect function it preserves, not the proof pattern.
The rewrite set is exactly six rules. (R1) Erase aliases and display names
to Catalog stable roles. (R2) Canonicalize qualified field, function, and
operator spellings and typed literals. (R3) Sort commutative components---AND
conjuncts, projections, aggregates, group keys, union operands, and join
operand/predicate order---by typed stable identity. (R4) Fold an accepted
connected equi-join graph into one canonical binary term (after the graph
boundary only, never reassociating arbitrary SQL). (R5) Collapse an
identical duplicate union branch only when its child carries a full-tuple
distinctness proof (Union and Group produce one, Select and Page preserve
it, Project preserves it only for an unchanged schema, Scan has none).
(R6) Canonicalize the Catalog page-order suffix. The supplement states a
per-rule effect-preservation lemma with these side conditions; RQ2's
zero-charge guarantee covers exactly R1--R6.

For a fixed approved task, let $\varpi_T$ be the normal-form profile bound in
$C$ and define the canonical replay descriptor
\begin{equation}
 \operatorname{RK}_T(q)=\langle P,S,C,\varpi_T,NF_{\Pi,\varpi_T}(q),
     \operatorname{params}_\Pi(q),\operatorname{encver}\rangle,
 \label{eq:replay-descriptor}
\end{equation}
where every component uses the typed, length-delimited encoding bound by the
Grant and Catalog; the implementation indexes a digest of this descriptor but
verifies the bound metadata before accepting a hit. Descriptor equality fixes
policy, publication, effect/View contracts, typed parameters, and normal form,
hence all three effect sets, so exact replay has zero novelty; the converse is
not claimed outside the stated rules. A semantic-View root compiles to a
closed core term with a binding $\beta_v$; DAG expansion retaining
$\beta_v$ preserves the visible bag and all three effect sets (proof in the
supplement); a direct terminal query can carry different Outcome FactIDs.

\begin{proposition}[Base-ledger split/page/join conservation]
\label{thm:base-conservation}
Merging release/dependency observations for disjoint projections or a complete
page partition equals observing the combined result. Overlapping pages and
repeated terminal observations charge each release/dependency FactID at most
once; join fanout changes witness multiplicity but not the union of dependency
support; predicate atoms are reused exactly when their canonical footprint
components agree (proof in the supplement; the support/multiplicity split is
the lineage projection of semiring provenance
\cite{green2007provenance,amsterdamer2011aggregate}).
\end{proposition}

\section{Online Accounting}
\label{sec:online}

Terms: a publication's \emph{hot artifact} holds its entity-key-to-ordinal
map and verified hashes, its \emph{cold} payloads the content-addressed Fact
bytes, and its \emph{sidecar} the PostgreSQL table mapping attested entity
keys to ordinals; the \emph{root head} is the Control row holding a family's
three set digests and epoch, advanced only by compare-and-swap (\emph{CAS}).

Formal semantics and the RQ1 oracle cover scan/select/project, group (with
HAVING), inner equijoin, union-distinct, and page. Online lowering
admits one product, a connected 2--16-source equi-join graph, or two-branch
union-distinct, optionally grouped; the SQL-AST path covers the first two and
the semantic-View path eliminates a bounded nested DAG into them. Union is
structured-plan only, pagination is one-product only, and ``online'' does
not mean arbitrary SQL. For joins, the PostgreSQL AST and Catalog types
feed the canonical $G_J$ builder, which emits the multi-way join operator
and the deterministic binary SQL/effect fold. Unsupported joins, set trees,
and pagination fail closed, and callers cannot supply roles, SQL fragments,
FactIDs, or provenance. For semantic Views, one catalog snapshot checks
definitions, interfaces, and dependency edges, expands safe acyclic Views
to materialized leaves, and emits the same core. Four digests (root
definition, dependency closure, canonical plan, public interface) bind into
the Grant and query; drift requires a new approval, and execution resolves
least-privilege terminal scopes before reservation.

\subsection{Snapshot Publication and Ordinal Compilation}

The publication compiler turns each frozen snapshot into immutable
per-row/per-field segments, sorting full Fact hashes and assigning 32-bit
ordinals (a FactID--ordinal bijection $D_j$ per segment; Roaring compression
changes representation, not membership \cite{chambi2016roaring}). A sidecar
maps entity keys to ordinals, payloads stay cold, and one manifest binds
snapshot, schemas, sidecar, dictionaries, and digests, so any collision
blocks publication.
Online, visible SQL and the companion run in one read-only
\code{REPEATABLE READ} transaction. The companion retains every required
dependency, and a k-way hash merge reproduces the witness commitment.
Release/Dependency use segmented Roaring bitmaps, and Outcome hashes enter an
exact Merkle-radix set. Invalid handles, manifests, encodings, overflow,
or truncation fail closed.

Decoding is exact: within one dictionary segment, bitmap OR, ANDNOT, and
popcount correspond to set union, difference, and cardinality, and for
every successful query in the closed online algebra the streamed bitmap
components decode to exactly the release and dependency components of
$\effect(q,D_s)$ (proof in the supplement).
For Outcome, hash prefixes route immutable content-addressed leaves and
blocks, but membership always compares the full SHA-256 value; leaf/block
encodings are supplemental.

\begin{proposition}[Exact incremental Outcome-set refinement]
\label{thm:outcome-radix-refinement}
Let $K_O$ be a valid Outcome root and $C_O$ a candidate. Loading every
leaf family whose 16-bit prefix occurs in $C_O$, performing exact full-hash
difference/union there, rebuilding only their blocks, and
reusing all other references yields
$\operatorname{Decode}(K'_O)=\operatorname{Decode}(K_O)\cup C_O$ and charges
$C_O\setminus\operatorname{Decode}(K_O)$ exactly. Any changed member, count,
prefix, chunk, block, or root changes a verified digest barring
SHA-256 collision.
\end{proposition}
\emph{Argument in the supplement} (prefix routing partitions the full-hash
universe; comparison is over all 256 bits; digests form a Merkle tree
\cite{merkle1987signature} over radix blocks).

\subsection{Atomic Ledger Settlement and Fail-Closed Artifact Publication}

For one successful query (Fig.~\ref{fig:architecture-settlement}b) the
enforcement layer reauthorizes, reserves resources, and looks up semantic
replay under the signed grant/publication binding. On a miss it executes
the visible and ordinal statements in one Business-PG snapshot, derives all
three effects, and stages encrypted Parquet (a hit loads a committed
observation; staging is never AVAILABLE). One Control-PG transaction then
computes exact bitmap/full-hash novelty against the root head, rejects any
$B_X$ excess, CASes all three manifests (conflicts reread and recompute
every dimension), and settles resources, observation, PENDING intent,
audit, and the signed receipt. A second transaction co-commits AVAILABLE
and \code{QUERY_RESULT_CONSUMED} after verifying promotion to the
canonical object, then returns the ID.

Only root heads are mutable: pre-settlement failure rolls back heads,
observations, intent, audits, cache, and receipt, while post-settlement
promotion failure neither refunds nor re-executes but resumes the charged
PENDING artifact idempotently after verifying the canonical digest;
$(\code{root_task_id},\code{observation_sha256})$ then identifies
zero-novelty replay without rescanning effects.

The boundaries remain distinct: \emph{accounted} means settlement committed,
\emph{available} means canonical promotion plus consumption audit, and
\emph{observed} means an untracked download. Thus
$\mathsf{Observed}\subseteq\mathsf{Available}\subseteq\mathsf{Accounted}$ and
$\mathsf{Available}\Rightarrow\mathsf{Settled}$, but not conversely.

Unlike request idempotence, semantic replay binds task/grant, typed normal
form/parameters, Catalog/schema/publication/dictionaries, and every profile
and encoding version; a hit skips Business/provenance SQL but still
reauthorizes, meters, audits, receipts, and re-encrypts; any mismatch takes
the novel path, and $P_{e+1}$ roots cannot hit $P_e$ entries.

\subsection{Receipts and Failure Semantics}

A signed Ed25519 settlement receipt binds accounting evidence, result/object
hashes, the \code{PENDING} intent, and registration audit without exposing
values or FactIDs; availability separately requires the hash-chained
\code{QUERY_RESULT_CONSUMED} inclusion, and a dedicated receipt domain
prevents cross-identity replay. Failures before connector invocation release
reservations, over-budget settlement creates no object or AVAILABLE row,
ambiguous connector outcomes charge resources but no exposure, and terminal
replay returns the first status without re-executing.

A rejected attempt returns only the code
\code{EXPOSURE_BUDGET_EXHAUSTED}---no dimension, count, or row (evaluation
tables name the oracle's projection, not a response field). It charges no exposure, in particular no Outcome atom, and
is recorded as a failed query (contributing no set,
Definition~\ref{def:task-exposure}) that consumes $B_Q$ and its admission
reservation (Section~\ref{sec:model}). $B_Q$ is an exposure parameter,
not a resource guard.

\begin{proposition}[Probing bound]
\label{prop:probing-bound}
Each rejected attempt reveals one comparison outcome---whether its novel
effect fits the remaining budgets---and consumes one unit of the attempting
task's own $B_Q$, which the prototype accounts per task. A delegated child
signs its own $B_Q$ no larger than its parent's, so a task tree of $k$ tasks
obtains at most $\sum_{i\le k}B_Q^{(i)}$ comparisons in total, independently
of $B_R$, $B_D$, and $B_O$. Each successful novel query additionally
returns its charge $|\Delta_X|$, the size of its footprint outside the
ledger; at most $B_O$ such counts exist per root family.
\end{proposition}
That a refusal itself answers a query is the classical auditing observation
\cite{denning1980tracker,kenthapadi2005simulatable,nabar2006robustness}.
The bound is on the response payload: execution and derivation precede
settlement (Section~\ref{sec:online}), so a refused query's latency may be
a function of the footprint that was not released, a timing channel the
model omits (Section~\ref{sec:limits}). The bound counts comparisons, not
what a comparison reveals about unreleased facts, and it grows with the
approved task tree because request
budgets, unlike exposure budgets, are not pooled at the root; charging
Outcome atoms on rejection would fold this capacity into $B_O$ at the cost of
penalizing legitimate near-boundary queries, and is not implemented. The
protocol is not a simulatable auditor and makes no denial-noninterference
claim.

\section{Delegation}
\label{sec:delegation}

\subsection{Root-Family Conservation}

Every child grant is a human-approved authorization (the prototype's only
approval mode is manual), so autonomous delegation is not supported and
conservation is evaluated for human-approved trees. A child authorization
signs both \code{root_task_id} and \code{parent_task_id}. Its human subject, catalog, datasource, schema, and
profile remain fixed, while products, columns, scopes, sensitivity, expiry, and
budgets may only narrow; every execution revalidates the signed
parent grant and ancestor chain; View-bound descendants inherit
the parent's contracts and cannot add or replace one.

All descendants resolve to one root ledger: if a parent and child request
the same fact, only the first successful settlement accounts it. Revoking
the root blocks new descendant queries while a terminal request ID stays
readable, and creating workers or task nodes cannot create fresh exposure
capacity. Every child grant is itself a human-approved authorization on
the Catalog's route (the prototype's only mode is manual approval), so the
tree's size, and the summed request budget of
Proposition~\ref{prop:probing-bound}, is set by approvals, not by the
agent. \emph{Example (cont.).} A delegated
child re-issuing the split $\pi_{id}$ replays at $0/0/0$ against the
shared ledger; a child attempting $\sigma_{dept='T'}$ is refused by the
same $\Delta_T$. Settlement checks the settling task's signed ceiling against the
family total, so a lower-ceiling child adds novel facts only while the
family total stays within it and otherwise only zero-novelty
replays: a narrower child cannot widen the family.

Conservation stops at the root and at the publication epoch. A second root
for the same principal, or a root approved on $P_{e+1}$, has an independent
ledger. Two approved roots on one Product thus double its capacity, and
unchanged content may be accounted again under $P_{e+1}$ (snapshot identity
is part of FactID identity). An existing root continues against $P_e$ through
a retained route or fails closed. Cross-root deduplication would need a
tenant-wide semantic identity outside the task-scoped guarantee.

Because Definition~\ref{def:task-exposure} sums over the queries that
settled, $K_X(T)=E_X(T,Q)$ is not itself a theorem. The claim is a safety
property of the following abstract settlement system. A root state is
$(K,e)$: three decoded sets and an epoch. A settlement transaction for
$q$ reads $(K,e)$, computes $\Delta_X=\facts_X(q,S)\setminus K_X$, and
attempts one compare-and-swap that publishes $(K\cup\facts(q,S),e+1)$ only if
the swap sees epoch $e$ and $|K_X|+|\Delta_X|\le B_X$ for all $X$; a swap
that sees $e'\neq e$ rereads; a rejected or failed attempt
publishes nothing. Root-family transactions, descendants included,
interleave arbitrarily but each swap is atomic.

\begin{property}[Atomic root-family budget safety]
\label{thm:root-family-safety}
In every execution of this system, every committed head equals the
componentwise union of the effects of exactly the queries whose swap
succeeded and satisfies $|K_X|\le B_X$ for all $X\in\{R,D,O\}$; the
property is unchanged when $K_X$ is represented by dictionary bitmaps or the
Merkle-radix set (Proposition~\ref{thm:outcome-radix-refinement} and the
supplement's bitmap-refinement proposition).
\end{property}
\emph{Argument.} Induct on successful swaps in linearization order
\cite{herlihy1990linearizability}: the
initial head is empty, each successful swap extends the head by exactly the
effect it computed against the head it read, and the bound check precedes
the swap; the \code{ExposureLedger} model (Section~\ref{sec:formal}) and
the Control-PG race and fault-injection tests are correspondence evidence,
not a mechanized refinement.

\subsection{Budget Calibration}
\label{sec:calibration}

\textbf{Catalog-issued capacity.} Table~\ref{tab:profile-calibration}
separates profile units from resource guards and from quantities they do
not estimate; unit weights enable exact set union, not harm or
differential-privacy accounting. These declarations are neither risk scores nor optimization targets:

\begin{table}[t]
\centering
\caption{Profile calibration boundary.}
\label{tab:profile-calibration}
\scriptsize
\setlength{\tabcolsep}{2.5pt}
\begin{tabular}{L{0.16\columnwidth}L{0.42\columnwidth}L{0.30\columnwidth}}
\toprule
Budget & Governs & Does not represent \\
\midrule
$B_R$ & Distinct accounted Result Fact identities selected for publication & Knowledge value or privacy loss \\
$B_D$ & Distinct profile-declared positive-output Dependency Facts & Minimal causes or all physical reads \\
$B_O$ & Distinct canonical predicate atoms and query/result composites & Bits, entropy, or DP $\epsilon$ \\
$B_Q$, rows, DB time & Request and resource availability & Semantic novelty or exposure deduplication \\
\bottomrule
\end{tabular}
\end{table}
 1,000
permits 1,000 declared units, and the bound profile must remain acceptable
when fully used. $B_O$ bounds distinct successful queries (every page
settles one composite) plus distinct predicate atoms: unlike the counters
Section~\ref{sec:introduction} dismisses it charges a repeated outcome once
and bounds no release size, which $B_R$ and $B_D$ do. Which dimension binds
depends on the purpose archetype. \emph{Detail review} (release named
rows): $B_R$, sized as permitted rows times fields, binds, and $B_D$ tracks
it. \emph{Single-Product aggregate reporting}: $B_D$ sized to the largest
admitted aggregate's footprint equals the Product's fact count, so within
that Product it degenerates and protects only against reading further
Products or columns; $B_O$ (expected pages plus distinct predicates) is
then the binding dimension, and the residual per-probe leakage is
Section~\ref{sec:online}'s quantified channel ($B_Q$ refusal bits, one
$|\Delta_X|$ per settled probe). \emph{Cross-Product join}: the joined
footprint exceeds any single Product's, so $B_D$ binds again. In every
archetype $B_Q$ is a small multiple of $B_O$; the accounting is compliance
accounting with quantified probe capacity, not extraction resistance. The
supplement works the campaign's bounded profile (its budgets and refusal
point: Section~\ref{sec:evaluation}), the count-probe ladder, and a page
sweep---a toy department, not a realistic-scale calibration.

\section{System Realization}
\label{sec:system}

The Go prototype uses separate Control and Business PostgreSQL 16 databases;
the Catalog declares Views, scopes, immutable publications, entity keys,
operations, approval routes, and profiles, and startup and per-query
attestation verify every binding. Each service instance is Catalog-bound,
so rollover routes old roots to retained $P_e$ instances and new approvals
to $P_{e+1}$. MCP tools cover discovery, task approval/execution/delegation,
View rebind, budgets, results, and receipts; direct SQL is lowered to the
same canonical plan or fails closed, so no path executes agent-authored
text; results use AES-GCM and audits form a hash chain. \system fronts the
governed serving tier (lake $\rightarrow$ curated products $\rightarrow$
immutable publications $\rightarrow$ \system $\rightarrow$ agent) and replaces
none of the surrounding lake, Catalog, IAM, OA/BPM, object-store, RLS/DFC,
or audit systems (supplement interface table; not a production deployment);
OA/BPM may impose cross-root issuance policies while the ledger stays
scoped to one root family. Crashes, concurrency, stale
bindings, replay, and promotion failure are in scope; compromised trusted
authorities, direct bypass, colluding independent roots, and external-tool
leakage are not.

\section{Three-Layer Assurance}
\label{sec:formal}

Paper theorems establish representation-independent Fact identity,
deterministic effects, and exact novelty for the closed algebra, and
Property~\ref{thm:root-family-safety} states root-family conservation over an
abstract settlement system; none proves the parser or implementation. Four
bounded TLC models \cite{lamport2002specifying} check abstract root
settlement, FactSet/bitmap correspondence, exact Outcome-set settlement with
collision fail-closed safety, and PENDING/AVAILABLE recovery at small
configurations (two facts per dimension, limits $1/1/2$;
\ExposureFormalStates{}, \BitmapFormalStates{}, \OutcomeFormalStates{},
and \ArtifactFormalStates{} states; bounds in the supplement): sanity
checks of the specification that abstract away SQL, rows, hashes,
signatures, and ciphertext. Executable Go/PostgreSQL differential, concurrency, tamper, and
fault-injection tests check implementation correspondence (the supplement
maps layers and gaps). Among them, a randomized root-family test settles
\VFiveFamilyPropertyQueries{} random observation sets over
\VFiveFamilyPropertySeeds{} random delegation trees (up to
\VFiveFamilyPropertyMaxTasks{} tasks, \VFiveFamilyPropertyNarrowedTasks{}
of them with ceilings drawn below their parent's) in sequential steps and
concurrent bursts against the live Control store. After every step it
checks that the committed head equals a separately kept plain-set union in
all three dimensions, never exceeds the root's or the settling task's
ceiling, and matches the ceiling rule's admit/refuse decision and
exact-novelty charge (\VFiveFamilyPropertyRefused{} refusals in all). It
drives the settlement engine directly, not SQL lowering. We claim bounded
model checking plus implementation evidence, not a mechanized refinement.

\section{Evaluation}
\label{sec:evaluation}

\subsection{Questions and Reproducibility}

RQ1 asks ground-truth Fact membership; RQ2, rewrite and charge invariance;
RQ3, conservation under adaptive execution, delegation, and concurrency; RQ4,
ordinal/bitmap acceptance (answered by the campaign's Scale profile; the
archived V4 run is reported only in the supplement); and RQ5, daily
publication and retained routing.
RQ1--RQ3 regenerate digest-bound corpus \ExposureCorpusHash{}, and every call
uses a complete signed profile. Medians and p95 are Type-7
quantiles \cite{hyndman1996quantiles} over the stated samples; ProvSQL
provenance capture is the same-host cost reference
(Section~\ref{sec:attack-provsql}).

RQ1's oracle (\RQOnePassed{}/\RQOneCases{} cases) is a separate
implementation over \RQOneRows{} rows that imports no production policy
code. The campaign's oracle package is likewise a separate implementation
of the FactID wire contract, written by us from the same specification. Its
import boundary excludes
every production package (enforced by a dependency test). It fixes the
expected Fact cardinalities of all
\FinalVFivePublicationIndependentOracleSamples{} Scale and trace samples
and is linked member by member, through the activated publication
dictionaries, to every committed Dependency set of the Scale and ProvSQL
cells (\FinalVFivePublicationScaleDependencyLinksVerified{} and
\FinalVFivePublicationProvSQLDependencyLinksVerified{} set verifications,
none with a missing or extra ordinal, up to
\FinalVFivePublicationScaleDependencyLinkMaxFacts{} Facts per set).
Agreement is thus implementation-to-specification, not a third party's
reading. Such oracles cannot detect specification errors or a shared
misreading of PostgreSQL semantics, so we claim consistency with our oracle
rather than independent validation; RQ1's \RQOneCases{} and RQ3's
\RQThreeCases{} cases are regression evidence, not statistical evidence.
One check shares nothing with our specification: a pilot-class rerun
(\FinalVFiveProvSQLLeafSamples{} samples) expands ProvSQL's own
circuit to its base tuples on the paired join cells and finds them equal,
member by member, to the oracle's \FinalVFiveProvSQLLeafTotalRowFacts{}
base-row facts, to which \system's Dependency sets are linked in turn.
The archived \RQFourVFourOracleFacts{}-Fact RQ4 check
(\RQFourVFourGatesPassed{}/\RQFourVFourGatesTotal{} gates, one index build)
shares the FactID encoder and is only derivation-independent. RQ1--RQ3
and RQ5 are regenerated at the submission commit; the campaign is sealed at
\FinalVFivePublicationCommit{}. RQ2
(\RQTwoExposureCases{} cases, \RQTwoExposureMismatches{} mismatches)
requires equal Result/Dependency sets and zero bidirectional charge. RQ3
(\RQThreeCases{} cases, deterministic and as PostgreSQL race tests) covers
split, page, retry, snapshot identity, joins/unions, and delegation.
RQ5 passes \RQFiveChecksPassed{}/\RQFiveChecksTotal{} rollover checks
across \RQFiveTransitions{} transitions (build and activation timings:
supplement). Outcome
regressions cover empty
probes, replay, radix tamper, and digest binding. Beyond the hand-written
cases, a generated-plan campaign (\GeneratedPlans{} random plans over
\GeneratedFixtures{} fixtures, spanning selections, projections,
equijoins, aggregates, and pages) compares the production algebra with a
separate reference evaluator: \GeneratedMismatches{} set and
\GeneratedHashMismatches{} hash mismatches over \GeneratedReleaseFacts{}
Result and \GeneratedDependencyFacts{} Dependency Facts, with split and
page unions equal to the complete query in all \GeneratedSplitChecks{}
checks (supplement).

The core workloads (the campaign's \emph{Baseline} profile): S1, a
two-column projection of the frozen 50,000-order corpus under an order-key
threshold (SF1: 5,000 orders; SF10: all); S2, the orders--lineitem equijoin
grouped to three rows; S3, a one-column projection over 1,000--45,000 rows
of the 414,000-row exposure-scale Product; S4, the depth-4 semantic-View
form of S2/SF1; S5, a two-branch union-distinct of S1's shape; S6,
result-heavy releases of 100--100,000 rows with 4 or 16 columns. Modes
are direct (the same SQL against PostgreSQL without \system), novel (the first
governed execution), and the replays; ``sealed campaign'' denotes the frozen
protocol's runs (retained pilots are supplementary, never eligible). The frozen
campaign (\FinalVFivePublicationCampaign{}, commit
\FinalVFivePublicationCommit{}) executed all
\FinalVFivePublicationTotalCells{} frozen cells:
\FinalVFivePublicationProfileCells{} profile cells across
\FinalVFivePublicationDeployments{} fresh deployments of
\FinalVFivePublicationProfiles{} profiles, each deployed
\FinalVFivePublicationFreshExecutions{} times, plus
\FinalVFivePublicationScaleNonProfileCells{} Scale and
\FinalVFivePublicationCompilerNonProfileCells{} Compiler deployment-free
cells. Every one of its \FinalVFivePublicationMeasuredSamples{} measured
samples passed (all per-sample invariant and oracle checks; deterministic,
not statistical), with a verified Receipt on every governed arm.
Compiler-admitted join widths and role counts beyond the measured cells
stay unmeasured (supplement).
Table~\ref{tab:baseline-main} gives
the per-cell medians: across the \FinalVFivePublicationBaselineCells{}
Baseline cells (sixteen workload/scale pairs with direct, novel, and the
protocol-defined replay cells) a novel governed query costs
\FinalVFivePublicationOverheadMin{}--\FinalVFivePublicationOverheadMax{}
times its direct PostgreSQL equivalent. That cost follows the total Fact
effect the query materializes, not its visible row count. S2/SF10 returns
\FinalVFivePublicationSTwoRows{} grouped rows over
\FinalVFivePublicationSTwoDependencyFacts{} dependencies in
\FinalVFivePublicationSTwoNovelS{}~s, longer than S1/SF10 needs to release
\FinalVFivePublicationSOneReleaseFacts{} Facts
(\FinalVFivePublicationSOneNovelS{}~s). The result-heavy S6 cells cost
more per Fact than S2, so Release materialization is at least as expensive
per Fact as Dependency derivation. Settlement is not where exactness is
paid: the Scale profile below settles against a million-Fact history in
\FinalVFivePublicationScaleSettleMillionMS{}~ms. A sub-phase pilot (\FinalVFiveSubphaseSamplesSTwoSFTen{} pilot sample per
cell; a supplementary pilot; supplement) splits execute-and-derive:
single-threaded Go-side ordinal consumption and witness merging dominate
(\FinalVFiveSubphaseGoSidePctSOneSFOne{}--\FinalVFiveSubphaseGoSidePctSSixHundredKByteSixteen\%
of the phase,
\FinalVFiveSubphaseMicrosPerFactSTwoSFTen{}--\FinalVFiveSubphaseMicrosPerFactSSixHundredKByteSixteen{}~$\mu$s
per Fact). Multi-second latency thus suits batch, not interactive, agents.
The bottleneck is single-threaded implementation, and replay buys database
silence, zero charge, and audit identity, not time. A retained cost
ablation (supplement) adds the accounting-off arm the sealed campaign lacks. An executed exposure-free Catalog
twin (authorization, publication, receipts, no fact accounting) costs
\BFiveSOneCommonPct\% of the S1/SF1 total and \BFiveSSixCommonPct\% of the
1.7-million-Fact S6 total. Fact derivation is the largest accounted share
(\BFiveSSixDerivation{} of \BFiveSSixArmIV{}~ms at S6), and the history-set
leaves take \BFiveSSixLedger{}~ms. On the sealed Scale Dependency sets a naive one-row-per-fact
exact ledger (unpaired pilot, supplement) takes
\NaiveScaleTenKFullRatio{}--\NaiveScaleTenKZeroRatio{}$\times$ \system's settlement time at $10^4$
facts and \NaiveScaleMillionFullRatio{}--\NaiveScaleMillionZeroRatio{}$\times$ at $10^6$.

\begin{table}[t]
\centering
\caption{Core workloads per cell (medians, ms, over
\FinalVFivePublicationBaselinePerCell{} samples): rows, Facts, direct, novel,
semantic and idempotent replay; overhead is novel/direct; --: no such cell.}
\label{tab:baseline-main}
\scriptsize
\setlength{\tabcolsep}{1.4pt}
\begin{tabular}{lrrrrrrrr}
\toprule
Cell & Rows & Rel. & Dep. & Direct & Novel & Sem. & Idem. & Over. \\
\midrule
\FinalVFivePublicationBaselineTableBody
\bottomrule
\end{tabular}
\end{table}
All \FinalVFivePublicationReplayZeroSQLSamples{} replay samples issued no
business SQL and charged no Facts (row budget still consumed). Replay is no
latency win: semantic replay exceeds direct re-execution in
every cell; only idempotent replay at S2/SF10 undercuts it.

The supplement's phase table shows where that time goes. The Gateway
reports six non-overlapping server phases per governed response. The
largest phase of S1/SF1 is \texttt{\FinalVFivePublicationPhaseDominantSOneSFOne{}}
(\FinalVFivePublicationPhaseDominantSharePctSOneSFOne\% of its server total) and
of S6/100k-x16 is \texttt{\FinalVFivePublicationPhaseDominantSSixHundredKByteSixteen{}}
(\FinalVFivePublicationPhaseDominantSharePctSSixHundredKByteSixteen\%).
The Control settlement transaction that computes exact novelty and advances
the root head takes \FinalVFivePublicationPhaseSettlementMSSOneSFOne{}~ms at
S1/SF1 and \FinalVFivePublicationPhaseSettlementMSSSixHundredKByteSixteen{}~ms
at S6/100k-x16. The other four phases (prepare, artifact staging,
publication, finalize) are the governed floor every response pays,
replays included; the semantic-replay median of a cell thus bounds that
floor, and the rest of the novel median is derivation plus settlement.
Same-root contention is measured separately: rounds of 10 and 50 concurrent
requests against one root family at its budget boundary. A single request
drains in \FinalVFivePublicationConcurrencySerialDrainPFiftyMS{}~ms (serial
control). A round of 50 under natural contention drains in
\FinalVFivePublicationConcurrencyFiftyNaturalDrainPFiftyMS{}~ms p50 and
\FinalVFivePublicationConcurrencyFiftyNaturalDrainPNinetyFiveMS{}~ms p95,
i.e., \FinalVFivePublicationConcurrencyFiftyNaturalRequestsPerSecond{}--\FinalVFivePublicationConcurrencyTenNaturalRequestsPerSecond{}
contending requests per second across the measured widths and modes through
one root head, one of which settles novel work per round. Every generated round met the contention condition (30
per cell and deployment, none discarded), and the supplement tabulates every
width and mode. Because every descendant settles on one root head, a task
tree's settlement throughput is bounded by that head; per-dimension or
batched settlement could relax the bound. With budget to spare, one head
settles \ThroughputOneRootFiftyDisjointNPS{} novel queries per second
(four roots \ThroughputFourRootFiftyDisjointNPS{}). At fifty contenders on
one head, \ThroughputOneRootFiftyRefused{} of
\ThroughputOneRootFiftyRequests{} requests returned \code{CONFLICT} after
sixteen head-CAS retries---uncharged, ledger exact, client resubmits---and
with up to \RetryK{} resubmissions \RetryOneRootFiftyCompletionPct\% settled
at an end-to-end p95 of \RetryOneRootFiftyDisjointEndToEndPNinetyFive{}~ms
(supplement). Table~\ref{tab:scale-main} answers RQ4 with the Scale profile
(\FinalVFivePublicationScaleCells{} cells,
\FinalVFivePublicationScaleSamples{} samples, every charge equal to the
oracle's union minus history): settlement against a same-size history is
sub-second at a million Facts and execute-and-derive dominates the response.

\begin{table}[t]
\centering
\caption{Scale profile (RQ4): settlement and full-response p50 (ms) for one
candidate against a same-size pre-seeded Dependency history.}
\label{tab:scale-main}
\scriptsize
\setlength{\tabcolsep}{4pt}
\begin{tabular}{lrrr}
\toprule
History (Facts) & Settle, 0\% ovl. & Settle, 100\% ovl. & Response \\
\midrule
$10^4$ & \FinalVFivePublicationScaleSettleTenKMS{} & -- & -- \\
\FinalVFivePublicationScaleMillionHistory{} & \FinalVFivePublicationScaleSettleMillionMS{} & \FinalVFivePublicationScaleSettleMillionFullOverlapMS{} & \FinalVFivePublicationScaleMillionDrainMS{} \\
\bottomrule
\end{tabular}
\end{table}

\subsection{Adaptive Extraction and Provenance-Capture Cost}
\label{sec:attack-provsql}

\textbf{Scripted decompositions.} The attack profile replays a frozen
corpus (\FinalVFivePublicationAttackCorpusDigest{}) of
\FinalVFivePublicationAttackSequences{} sequences that try to obtain
governed Facts twice (pages and an overlap around a complete query in both
orders, four spellings of one predicate, a range split and its union,
replays, a count-probe ladder with delegation;
\FinalVFivePublicationAttackSamples{} samples, identical per-step charges
per cell). Each decomposition was charged its complete query's Facts
exactly once
(\FinalVFivePublicationAttackPaginationReleaseFacts{}/\FinalVFivePublicationAttackPaginationDependencyFacts{}
for the six-step pagination in either order), each distinct page or range
shape settled one composite, replays and rewrites charged nothing, and both
set-operation variants failed closed. In the ladder, two count probes (one
from a delegated child) and a detail primer consume the root's Outcome
ceiling of \FinalVFivePublicationAttackThresholdCeiling{}. The child's third
probe is refused with \FinalVFivePublicationAttackThresholdRejectionCode{},
so the agent learned \FinalVFivePublicationAttackThresholdAnswered{} of
\FinalVFivePublicationAttackThresholdProbes{} thresholds plus one refusal
bit while the direct arm answered all three (supplement). The corpus is
preregistered, and the adversary study above probes the strategy space.
A simulated eight-point recipe sweep (supplement) first admits all
\SweepBenignAuthorized{} authorized statements and first recovers the secret at one grid point, $\SweepRecoverAt{}\times$.

\textbf{Adaptive trace against per-query control.} The RLS profile replays
one deterministic \FinalVFivePublicationRLSQueries{}-query adaptive trace
(receipt look-ups, pages, equivalent predicate spellings, repeated
aggregations, branches chosen from prior scalars) through three arms on one
Product. The arms are PostgreSQL row-level security with a department policy and column
grants, \system with an unlimited ledger, and \system bounded at
$\lfloor 0.7\,U\rfloor$ of the trace's distinct facts
$U=\FinalVFivePublicationRLSUnionRelease{}/\FinalVFivePublicationRLSUnionDependency{}/\FinalVFivePublicationRLSUnionOutcome{}$
as counted by a separate trace-union oracle (budgets
\FinalVFivePublicationRLSBudgetRelease{}/\FinalVFivePublicationRLSBudgetDependency{}/\FinalVFivePublicationRLSBudgetOutcome{};
\FinalVFivePublicationRLSPerCell{} samples per arm, identical in every one).
RLS answered all \FinalVFivePublicationRLSQueries{} queries
(\FinalVFivePublicationRLSRowsRLS{} rows) in
\FinalVFivePublicationRLSDrainRLSMS{}~ms. \system-unlimited answered all in
\FinalVFivePublicationRLSDrainUnlimitedS{}~s, and its final ledger equals the
oracle's union. \system-bounded answered
\FinalVFivePublicationRLSBoundedSuccessful{} and
refused query \FinalVFivePublicationRLSBoundedRefusalIndex{}---the first
whose oracle prefix exceeds a budget (Fig.~\ref{fig:rls-trace})---with
\FinalVFivePublicationAttackThresholdRejectionCode{}; it released nothing
afterwards and ended at \FinalVFivePublicationRLSBoundedLedger{}. A control
step outside the policy rows and the column grant is refused at the column
under both (supplement), so \system composes with per-query authorization.
That RLS admits every query is definitional (no cumulative budget); the
comparison shows where a set budget binds. The counter comparators are also \emph{executed}, across three fresh
deployments that each realized the same outcome (retained pilot,
supplement). The same trace ran under row-, query-, and release-set budgets
fixed a priori by the same $\lfloor 0.7\,U\rfloor$ recipe per unit, in
three orders each (unlike the supplement's campaign-derived counter replay,
calibrated to the set ledger's stopping point). The counters
released the full \FinalVFiveCounterNaiveMinDep{}-fact Dependency union in
every order; the exact floors capped release at
\FinalVFiveCounterExactMaxDep{} in every order (supplement table). Executed arms compare default products
(counters truncate and archive; exposure refuses without archiving).
Replayed under one refuse-whole, task-alive rule, only the row counter
moves (\AlignedRowsNaturalExecuted{}$\to$\AlignedRowsNaturalAdmitted{}
admitted), and the set floors remain the only budget bounding the released
set (supplement).

\begin{figure}[!t]
\centering
\begin{tikzpicture}
\begin{axis}[width=0.98\columnwidth, height=0.5\columnwidth, font=\scriptsize,
  xlabel={query index in the adaptive trace}, ylabel={distinct facts (oracle)},
  xmin=1, xmax=100, ymin=0, ymax=34,
  ytick={0,\FinalVFivePublicationRLSBudgetRelease,\FinalVFivePublicationRLSBudgetDependency,\FinalVFivePublicationRLSBudgetOutcome,30},
  legend pos=north west, legend columns=3,
  legend style={font=\scriptsize, draw=none, fill=none}, grid=major, grid style={gray!25}]
\addplot[blue, thick, mark=o, mark repeat=10, mark size=1.2pt] coordinates {\FinalVFivePublicationRLSPrefixRelease};
\addplot[red, thick, mark=square*, mark repeat=10, mark size=1.1pt] coordinates {\FinalVFivePublicationRLSPrefixDependency};
\addplot[black!60!green, thick, mark=triangle*, mark repeat=10, mark size=1.4pt] coordinates {\FinalVFivePublicationRLSPrefixOutcome};
\legend{Release, Dependency, Outcome}
\addplot[blue, dashed] coordinates {(1,\FinalVFivePublicationRLSBudgetRelease) (100,\FinalVFivePublicationRLSBudgetRelease)};
\addplot[red, dashed] coordinates {(1,\FinalVFivePublicationRLSBudgetDependency) (100,\FinalVFivePublicationRLSBudgetDependency)};
\addplot[black!60!green, dashed] coordinates {(1,\FinalVFivePublicationRLSBudgetOutcome) (100,\FinalVFivePublicationRLSBudgetOutcome)};
\draw[gray, thick] (axis cs:\FinalVFivePublicationRLSBoundedRefusalIndex,0) -- (axis cs:\FinalVFivePublicationRLSBoundedRefusalIndex,34)
  node[pos=0.07, right, font=\scriptsize] {budget hit: bounded arm refuses};
\end{axis}
\end{tikzpicture}
\caption{Distinct governed facts by query index (trace-union oracle; solid)
against the bounded arm's budgets (dashed); the bounded arm refuses at the
first crossing.}
\label{fig:rls-trace}
\end{figure}

\textbf{Provenance-capture cost.} The ProvSQL profile (supplement) pairs
byte-identical grouped SQL on pinned PostgreSQL~16.14, ProvSQL~1.11.0, and
\system's full governed path: at 45k orders
\FinalVFivePublicationProvSQLFortyFiveKPostgreSQLMS{}~ms,
\FinalVFivePublicationProvSQLFortyFiveKProvSQLS{}~s, and
\FinalVFivePublicationProvSQLFortyFiveKBDGS{}~s over
\FinalVFivePublicationProvSQLFortyFiveKDependencyFacts{} Dependency Facts,
a boundary that is asymmetric by construction and ranks nothing.

\subsection{Reproducibility Boundary}

All measurements ran on one laptop-class WSL2 host with a digest-pinned
PostgreSQL 16.14 image (supplement). The artifact is the
repository at the cited commits (\url{https://github.com/minmin-lab/factgate},
with the evidence pack and the digest of every deposited file) and the
retained raw evidence (\RawEvidenceFiles{} files, \RawEvidenceMB{}~MB: the
campaign's \RawEvidenceSealedSamples{} sample files, its digest-chain records,
and the retained pilots' samples; Zenodo, doi:\RawEvidenceDOI{}).
Every digest chain and macro is recomputable
from it, as a fresh checkout confirmed. Numbers attributed to the campaign come from
the sealed campaign: one submission commit (\FinalVFivePublicationCommit{}),
\FinalVFivePublicationDeployments{} fresh isolated deployments on one host,
three executions per profile whose per-execution novel medians differ by at
most \FinalVFivePublicationNovelSpreadMaxPct\% of the pooled median (direct
\FinalVFivePublicationDirectSpreadMaxPct\%, concurrency
\FinalVFivePublicationConcurrencySpreadMaxPct\%). Numbers marked pilot,
retained, or simulated come from later non-publication runs; the supplement
maps each number to its class, run, and file. RQ1--RQ3 use small
fixtures; ordinal/bitmap evidence is warm, single-host, TPC-H-derived (not
official); RQ5 sums three sequential warm child intervals. No result is generalized to cold-cache, multi-node,
or production-SLO behavior.

\section{Limitations and Threats to Validity}
\label{sec:limits}

\textbf{Semantic and subject scope.} Outcome atoms and composites are
governance units, not bits; \system omits rejected rows, non-output
groups/pages, rank-only inputs, correlations, background knowledge, and
physical reads, and provides neither minimal causality nor differential
privacy (Section~\ref{sec:model}). \textbf{Query and identity scope.} Evidence uses
frozen, manifest-proven publications. The closed fragment is a
reporting-Product fragment, not ad-hoc analytics: the production lowerer
admits \TPCHLowerableExplicitJoin{} of the \TPCHQueries{} TPC-H templates
(first rejections: subqueries
\TPCHSubqueryQueries{}, arithmetic \TPCHArithmeticQueries{}, multi-product
\code{LIMIT}) and
\SSBLowerableExplicitJoin{} of the \SSBQueries{} Star Schema Benchmark
reporting queries (supplement tables).
For LLM-written SQL from the Product sheet alone, \HeldoutLowerable{}
(\HeldoutLowerablePct\%) of \HeldoutQueries{} held-out
questions written after the fragment froze lower unedited (admission, answers ungraded;
\SheetCorrectedRange{} under a corrected sheet, pilot). Of the rest,
\HeldoutEngineeringGaps{} are engineering
gaps, \HeldoutNewRule{} need new rules, and \HeldoutAgentDefects{} is agent output; they stop
first at scalar functions, division or a subquery inside
arithmetic, non-literal comparisons, \code{BETWEEN}/\code{LIKE}, \code{OR},
\code{LEFT JOIN}, or \code{LIMIT} over a join. Non-literal comparisons,
\code{OR}, and outer joins need new rules (sketched in the supplement). The
development set, adapted after its first pass, lowers
\AgentWorkloadLowerable{} of \AgentWorkloadQueries{}
(\AgentWorkloadLowerablePct\%) and measures iteration, not the frozen
boundary. In a two-objective agent study (\AgentRuns{} runs, supplement) an
off-the-shelf agent completes the legitimate objective in every run after one
refused listing, and its count-only probe stops at six probes. Stable keys and
publications are trusted declarations, with View drift checked at request,
approval, and execution.

\textbf{Applicability envelope.} Publication artifacts cost
\RQFiveBytesPerFact{}~B and \RQFiveBuildMicrosPerFact{}~$\mu$s per Fact
(RQ5 Day~0). A $10^9$-Fact Product
would need hundreds of gigabytes and hours per build, so the envelope is $10^5$--$10^6$-row Products; a
retained pilot ladder settles \FinalVFiveScaleSevenLargestDependency{}
Dependency facts in one query (supplement). Same-root
settlement serializes on the head (Section~\ref{sec:evaluation}).

\textbf{Governance and deployment scope.} Profile correctness is
trusted---\system validates bindings, semantics, and settlement, not policy
appropriateness---and the single database path and demo approval adapter
offer no production integration study \cite{tonnarelli2026dataproductmcp}. The guarantee composes with controls it
does not supply: task-tree and refusal-rate caps (Proposition~\ref{prop:probing-bound}),
timing padding, and root expiry; splitting one objective across
independently approved roots is fresh capacity no ledger spans, so its
defence is each root's approval, not accounting.

\section{Conclusion}

\system provides profile-relative exact accounting of Result, Dependency,
and Outcome sets within one root family behind one scoped claim: no
governed artifact becomes \code{AVAILABLE} before atomic settlement.

\bibliographystyle{IEEEtran}
\bibliography{references}

\clearpage
\onecolumn
% Supplementary material (compiled separately from supplement.tex in the
% repository; included here so the arXiv record carries both parts).
\includepdf[pages=-]{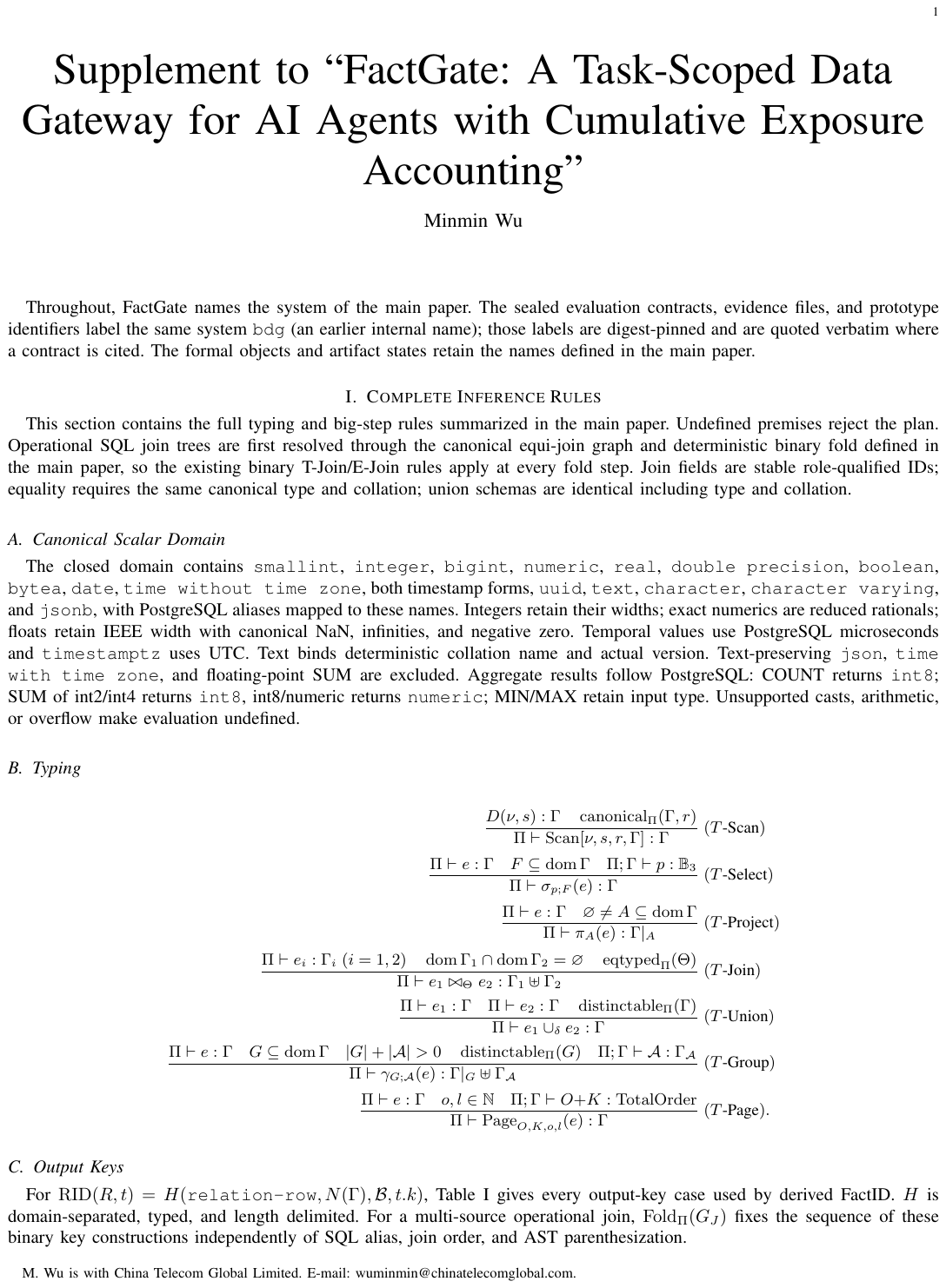}
\end{document}